\documentclass{article}
\usepackage[margin=1in]{geometry}
\usepackage[affil-it]{authblk}
\usepackage{natbib}
\usepackage{amsmath}
\usepackage{amsfonts}
\usepackage{bm}
\usepackage[color=lightgray]{todonotes}
\usepackage[bf,font={small,sl}]{caption}
\usepackage{booktabs}

\title{\textbf{Inference in MCMC step selection models}}
\author
{Th\'eo Michelot$^1$\footnote{tmichelot1@sheffield.ac.uk},
Paul G. Blackwell$^1$,
Simon Chamaill\'e-Jammes$^2$,
Jason Matthiopoulos$^3$ \\
$^{1}$ University of Sheffield, Hounsfield Road, Sheffield S37RH, U.K. \\
$^{2}$ CEFE-CNRS, 1919 route de Mende, 34293 Montpellier Cedex 5, France \\
$^{3}$ University of Glasgow, 82 Hillhead St, Glasgow G12 8QQ, U.K.}
\date{}

\begin{document}
\maketitle

\begin{abstract}
\noindent
\emph{Habitat selection models are used in ecology to link the distribution of animals to environmental covariates, and identify habitats that are important for conservation. The most widely used models of this type, resource selection functions, assume independence between the observed locations of an animal. This is unrealistic when location data display spatio-temporal autocorrelation. Alternatively, step selection functions embed habitat selection in a model of animal movement, to account for the autocorrelation. However, inferences from step selection functions depend on the movement model, and they cannot readily be used to predict long-term space use. We recently suggested that a Markov chain Monte Carlo (MCMC) algorithm could define a step selection model with an explicit stationary distribution: the target distribution. Here, we explain how the likelihood of a MCMC step selection model is derived, and how maximum likelihood estimation can be used for inference about parameters of movement and habitat selection. We describe the local Gibbs sampler, a rejection-free MCMC scheme designed to capture important features of real animal movement. The sampler can be used as the basis for a flexible class of movement models, and we derive the likelihood function for several important special cases. In a simulation study, we verify that maximum likelihood estimation can be used to recover all model parameters. We illustrate the application of the method with data from a plains zebra.}
\end{abstract}

\vspace{1em}
 \vspace{1em}

\section{Introduction}
Location data are routinely collected on individual animals, e.g.\ with GPS tags, resulting in bivariate time series of coordinates. Statistical methods have been developed to combine location data with environmental data, to understand how an animal's use of space relates to the distributions of some spatial covariates \citep[e.g.\ density of a resource, or habitat type;][]{manly2007resource}. The focus of such analyses is habitat selection, i.e.\ the lack of proportionality between habitat availability and habitat use by the animal \citep{northrup2013practical}. Resource selection functions (RSFs) are a popular tool to link animal location data to environmental covariates. A RSF $w(\bm{c})$ associates a vector of covariates $\bm{c} = (c_1,\dots,c_n)'$ to the relative probability of the animal using a spatial point with these habitat characteristics \citep{boyce1999relating}. It is usually formulated as a log-linear regression model,
\begin{equation} \label{eqn:rsf}
  w(\bm{c}) = \exp \left( \sum_{i=1}^n \beta_i c_i \right).
\end{equation}
The resource selection coefficients $\beta_i$ can be estimated from location data, e.g.\ using a Poisson GLM for binned data, or a logistic regression for use-availability data \citep{aarts2012comparative}. The RSF is used to model the long-term distribution of the animal's locations, termed the utilisation distribution,
\begin{equation} \label{eqn:ud}
	\pi(\bm{x}) = \dfrac{w(\bm{c}(\bm{x}))}{\int_\Omega w(\bm{c}(\bm{y})) d\bm{y}}
\end{equation}
where $\bm{c}(\bm{x}) = (c_1(\bm{x}),c_2(\bm{x}),\dots,c_n(\bm{x}))'$ associates a spatial location $\bm{x}$ to a vector of $n$ covariate values, and $\Omega$ denotes the study region. The regression parameters $\beta_i$ then reflect the effect of the covariates on the animal's use of space (i.e.\ preference or avoidance).

RSF models are usually based on the assumption that the observed locations are an independent sample from the utilisation distribution. They often ignore the autocorrelation inherent in animal telemetry data, or treat it as a nuisance \citep{fieberg2010correlation}. Additionally, to define habitat availability, RSF analyses typically assume that any location within the study area (e.g.\ home range, population range) is equally accessible to the individual \citep{matthiopoulos2003use, aarts2008estimating}.

Step selection functions (SSFs) offer an alternative framework, developed to address these limitations of RSFs \citep{fortin2005wolves, forester2009accounting, thurfjell2014applications}. In SSF models, the autocorrelation of tracking data is explicitly accounted for, with a joint model of animal movement and habitat preference. Habitat availability is specified by the movement model: given the current location of the animal, what spatial units are accessible to it within one time step? \citep{johnson2008general} An SSF model describes the likelihood of an animal moving from one location to another, given the corresponding covariate values:
\begin{equation} \label{eqn:ssf}
	p(\bm{y} \vert \bm{x}) = \dfrac{\phi(\bm{y} \vert \bm{x}) w(\bm{c}(\bm{y}))}
    {\int_\Omega \phi(\bm{z} \vert \bm{x}) w(\bm{c}(\bm{z})) d\bm{z}},
\end{equation}
where $\phi(\bm{y} \vert \bm{x})$ is the likelihood of a step from $\bm{x}$ to $\bm{y}$ in the absence of covariate effects, and $w$ measures habitat preference, and takes the form given in Equation \ref{eqn:rsf}. Matched conditional logistic regression is typically used to estimate the parameters of a step selection model from telemetry data \citep{fortin2005wolves}. The function $\phi$ is called the resource-independent movement kernel, and it characterizes the underlying movement model.

Although the same notation is often used in resource selection and step selection analyses, the methods do not lead to the same estimates for the regression coefficients $\beta_i$, and thus for the function $w$. In resource selection models, the coefficients are directly linked to the global distribution $\pi(\bm{x})$ of space use through Equation \ref{eqn:ud}. However, the coefficients of an SSF measure local habitat preference: their interpretation is tied to the choice of the movement kernel $\phi$. The RSF is a limiting case of the SSF, where $\phi$ is uniform over the whole study region $\Omega$. This corresponds to the animal being able to move to any point in space at every time step. The discrepancy between the approaches has been demonstrated analytically \citep{moorcroft2008mechanistic, barnett2008analytic}, and empirically \citep{signer2017estimating}. There have been efforts to develop methods to derive utilisation estimates from SSF models. In particular, \cite{potts2014predicting} described an approach to obtain a numerical approximation of the steady-state (utilisation) distribution of the SSF model given by Equation \ref{eqn:ssf}, based on a ``master equation''. Alternatively, \cite{avgar2016integrated} and \cite{signer2017estimating} suggested using simulations from a fitted SSF model to estimate its stationary distribution. Although their approaches offer a way to numerically evaluate the equilibrium distribution of an SSF model, that distribution cannot be written as a simple function of the spatial covariates (as in Equation \ref{eqn:rsf}).

\cite{michelot2018linking} introduced a new model of step selection, in which the steady-state distribution of the movement process is (proportional to) the RSF given in Equation \ref{eqn:rsf}. The central idea is to use a Markov chain Monte Carlo (MCMC) algorithm -- with target distribution the normalized RSF -- to define a model of animal movement. Here, we extend the approach of \cite{michelot2018linking} and show how it can be used to estimate habitat preference and movement characteristics from animal tracking data. We focus on a specific MCMC algorithm, the local Gibbs sampler, designed to capture essential features of animal movement. We present a simulation study to investigate the performance of the method, for a known utilisation distribution. Finally, we use the local Gibbs movement model to analyse the trajectory of a plains zebra (\emph{Equus quagga}).

\section{Markov chain Monte Carlo as a model of animal movement}
\label{sec:model}
To reconcile the RSF and SSF approaches to modelling animal movement and space use, we consider a model based on an analogy between the movements of an MCMC sampler in its parameter space, and the movements of an animal in geographical space, introduced by \cite{michelot2018linking}. 

\subsection{MCMC step selection model}
\label{sec:model1}
By construction, an MCMC algorithm describes step selection rules, determined by the transition kernel $p(\bm{x}_{t+1} \vert \bm{x}_t)$, such that the long-term distribution of samples $\{ \bm{x}_1, \bm{x}_2, \dots \}$ is a given distribution, termed the target distribution \citep{gilks1995markov}. As such, it can be considered as the basis for a model of animal movement: the transition kernel defines the movement rules of the animal, and the target distribution is the utilisation distribution $\pi$ (i.e.\ the long-term distribution of the animal's space use). To link the animal's movement to the distribution of the covariates of interest, we model the utilisation distribution with a (normalized) resource selection function, as given in Equation \ref{eqn:ud}. The tuning parameters of the MCMC algorithm here become of interest: they are the parameters of the movement process. These movement parameters, and the parameters of the target distribution (the $\beta_i$ in Equation \ref{eqn:rsf}), are unknown, and may be estimated from animal movement and habitat data. The resulting model describes an animal's movement in response to its environment, similarly to SSF models, and we call it an ``MCMC step selection model''.

In RSF analyses, the used locations arise independently from a spatial point process, and the aim of the study is to model its steady-state distribution. On the other hand, SSFs treat the locations as the output of a (possibly second-order) Markov process, and the focus is on its transition kernel. We combine the two approaches: in an MCMC step selection model, both the utilisation distribution of the animal and its movement density are modelled explicitly.

The many different algorithms within the MCMC family differ in the way the samples are generated, as described by their transition kernel $p(\bm{x}_{t+1} \vert \bm{x}_t)$. In the context of animal movement, each algorithm may capture some specific characteristics of the movement, and the choice of the MCMC sampler determines the choice of a movement model. Standard MCMC algorithms are based on discrete-time Markov chains, and the transition density is defined over a fixed (implicit) time step, corresponding to one iteration of the algorithm. In the analogy presented here, one MCMC iteration represents one time step for the animal. In most cases, including the local Gibbs model described in Section \ref{sec:model2}, an MCMC step selection model is thus defined in discrete time, and may only be used for the analysis of telemetry data collected at regular time intervals. 

Some MCMC algorithms may not provide a realistic description of animal movement, if the transition kernel is a poor representation of the animal's step selection rules. Here, we describe a generalised version of the local Gibbs sampler, introduced by \cite{michelot2018linking}, as the basis for an MCMC step selection model. 

\subsection{The local Gibbs sampler}
\label{sec:model2}
We extend the algorithm introduced by \cite{michelot2018linking}, to a more flexible family of step selection models. The sampler that they described is a special case of the algorithm presented in this paper, but we keep the ``local Gibbs'' name that they coined. We restrict our attention to the case where the local Gibbs algorithm is used to sample from a two-dimensional target distribution $\pi$ on $\Omega \subset \mathbb{R}^2$, as animal space use is generally described in the two-dimensional coordinate space. It is however straightforward to generalise it to higher-dimension spaces.

We choose a radially symmetric distribution with density $\phi(\cdot \vert \bm{x})$, centred on $\bm{x}$, from which we can generate samples. Starting from a point $\bm{x}_1$, the local Gibbs algorithm for the target distribution $\pi$ is defined by the following steps.

For $t=1,2,\dots$,
\begin{enumerate}
\item Generate a point $\bm\mu$ from $\phi(\cdot \vert \bm{x}_t)$.
\item Define the distribution $\tilde\pi$ on $\Omega$ by
\begin{equation*}
	\tilde{\pi}(\bm{x}) = \dfrac{\phi(\bm{x} \vert \bm\mu) \pi(\bm{x})}
    	{\int_{\bm{z} \in \Omega} \phi(\bm{z} \vert \bm\mu) \pi(\bm{z}) d\bm{z}}.
\end{equation*}
\item Sample $\bm{x}_{t+1}$ from $\tilde{\pi}$.
\end{enumerate}

The sampled points $\{ \bm{x}_1, \bm{x}_2, \dots \}$ have $\pi$ as their stationary distribution. Indeed, we can verify that this algorithm satisfies the detailed balance condition. We have
\begin{align*}
	\pi(\bm{x}_t) p(\bm{x}_{t+1} \vert \bm{x}_t) & = \pi(\bm{x}_t) \int_{\bm\mu \in \Omega} 
    	p(\bm{x}_{t+1} \vert \bm\mu) p(\bm\mu \vert \bm{x}_t) d\bm\mu \\
	& = \pi(\bm{x}_t) \int_{\bm\mu \in \Omega} \dfrac{\pi(\bm{x}_{t+1}) 
    	\phi(\bm{x}_{t+1} \vert \bm\mu)}
    	{\int_{\bm{z} \in \Omega} \pi(\bm{z}) \phi(\bm{z} \vert \bm\mu) d\bm{z}} 
        \phi(\bm\mu \vert \bm{x}_t) d\bm\mu\\
   	& = \pi(\bm{x}_t) \pi(\bm{x}_{t+1}) \int_{\bm\mu \in \Omega} 
    	\dfrac{\phi(\bm{x}_{t+1} \vert \bm\mu) \phi(\bm\mu \vert \bm{x}_t)}
    	{\int_{\bm{z} \in \Omega} \pi(\bm{z}) \phi(\bm{z} \vert \bm\mu) d\bm{z}} d\bm\mu\\
   	& = \pi(\bm{x}_{t+1}) \pi(\bm{x}_t) \int_{\bm\mu \in \Omega} 
    	\dfrac{\phi(\bm\mu \vert \bm{x}_{t+1}) \phi(\bm{x}_t \vert \bm\mu)}
    	{\int_{\bm{z} \in \Omega} \pi(\bm{z}) \phi(\bm{z} \vert \bm\mu) d\bm{z}} d\bm\mu\\
   	& = \pi(\bm{x}_{t+1}) p(\bm{x}_t \vert \bm{x}_{t+1}),
\end{align*}
as required. The local Gibbs sampler is thus a valid MCMC algorithm for any symmetric density $\phi$, with target distribution $\pi$. Note that it is a rejection-free sampler, as it does not need an acceptance-rejection step to preserve the correct equilibrium distribution.

Within the framework described in Section \ref{sec:model1}, the local Gibbs sampler defines a model of animal movement and habitat selection. The choice of the transition density $\phi$ determines the shape of the movement kernel. In the following, we consider the case where $\phi$ is a parametric density function, and we explicitly denote it $\phi(\cdot \vert \bm{x}, \bm\theta)$, where $\bm\theta$ is a vector of movement parameters. We discuss two special cases of $\phi$ in Sections \ref{sec:model3} (normal transition density) and \ref{sec:model4} (availability radius model). The intermediate point $\bm\mu$ sampled in step 1 of the local Gibbs algorithm does not have a biological interpretation, but it is necessary to define a valid algorithm. 

In the general case, the step density of the model from $\bm{x}$ to $\bm{y}$ is given by the transition kernel of the algorithm,
\begin{equation}
	\label{eqn:stepdens}
	p(\bm{y} \vert \bm{x}, \bm\theta) = \pi(\bm{y}) \int_{\bm\mu \in \Omega} 
    	\dfrac{\phi(\bm{y} \vert \bm\mu, \bm\theta) \phi(\bm\mu \vert \bm{x}, \bm\theta)}
        {\int_{\bm{z} \in \Omega} \pi(\bm{z}) \phi(\bm{z} \vert \bm\mu, \bm\theta) d\bm{z}} d\bm\mu.
\end{equation}
The steps of the derivation are similar to the proof of the detailed balance condition, given above. In the absence of covariate effects (i.e.\ if $\forall \bm{x} \in \Omega, \pi(\bm{x}) = k$), each step is the sum of two $\phi$-distributed increments. Thus the habitat-independent movement kernel is given by the convolution
\begin{equation*}
	p_0(\bm{y} \vert \bm{x}, \bm\theta) = \int_{\bm\mu \in \Omega} 
    	\phi(\bm{y} \vert \bm\mu, \bm\theta) \phi(\bm\mu \vert \bm{x}, \bm\theta) d\bm\mu.
\end{equation*}

In step 2 of the algorithm given above, the integral $\int_{\bm{z} \in \Omega} \phi(\bm{z} \vert \bm\mu) \pi(\bm{z}) d\bm{z}$ cannot generally be evaluated analytically, unless the covariates follow a tractable parametric form. In practice, at each iteration, a large number of points $\{ \bm{z}_1, \bm{z}_2, \dots, \bm{z}_K \}$ is sampled from $\phi(\cdot \vert \bm\mu)$, and $\bm{x}_{t+1}$ is selected from the $\bm{z}_i$, with probabilities given by $\pi(\bm{z}_i)/\sum_j \pi(\bm{z}_j)$. With this procedure, we can simulate movement tracks on a given target (utilisation) distribution; see Section \ref{sec:sim} for an example. The local Gibbs algorithm would usually not be an interesting choice for the general purpose of sampling from a probability distribution (e.g.\ in the context of Bayesian inference). Indeed, although there are no rejections, the numerical integration requires many evaluations of the target distribution for each sampled point, which renders the procedure more computationally intensive than, say, standard Metropolis-Hastings sampling. In the following, we only consider the local Gibbs algorithm for the purpose of modelling animal movement.

The local Gibbs algorithm is not limited to continuous distributions for the choice of $\phi$. In particular, it would be possible to define $\phi$ as the combination of a continuous symmetric distribution, and some probability mass at the origin. This model would allow for steps of length zero with positive probability, corresponding to time steps over which the animal does not move.

An extension of the local Gibbs algorithm can be obtained by considering that the parameters $\bm\theta$ of the transition density $\phi$ are themselves random, and are drawn at each iteration from a probability distribution $p(\bm\theta \vert \bm\omega)$. This results in a hierarchical model, formulated in terms of the hyperparameters $\bm\omega$. In this case, the step density becomes
\begin{align}
	p(\bm{y} \vert \bm{x}, \bm\omega) & = \int_{\bm\theta} 
		p(\bm{y} \vert \bm{x}, \bm\theta) p(\bm\theta \vert \bm\omega) d\bm\theta \nonumber \\
        & = \pi(\bm{y}) \int_{\bm\theta} p(\bm\theta \vert \bm\omega) \int_{\bm\mu \in \Omega} 
    	\dfrac{\phi(\bm{y} \vert \bm\mu, \bm\theta) \phi(\bm\mu \vert \bm{x}, \bm\theta)}
        {\int_{\bm{z} \in \Omega} \pi(\bm{z}) \phi(\bm{z} \vert \bm\mu, \bm\theta) d\bm{z}} d\bm\mu d\bm\theta.
   	\label{eqn:stepdens2}
\end{align}
This extension provides additional flexibility in the shape of the transition density, to define a more realistic movement model. We explore two important special cases of the local Gibbs model in Sections \ref{sec:model3} and \ref{sec:model4}; in the latter, we discuss the possibility of modelling the movement parameter as random.

\subsection{Normal kernel model}
\label{sec:model3}
An interesting special case of the local Gibbs model is obtained when $\phi$ is taken to be a bivariate (circular) normal density centred on the origin $\bm{x}_t$. In the absence of covariate effects (i.e.\ if the target distribution $\pi$ is flat), each displacement from $\bm{x}_t$ to $\bm{x}_{t+1}$ is the sum of two bivariate normal steps, which is also normally distributed. That is, if $\phi$ is a normal distribution with variance $\sigma^2\bm{I}_2$, where $\bm{I}_2$ is the $2 \times 2$ identity matrix, then the resource-independent movement kernel is also a normal distribution, with variance $2\sigma^2\bm{I}_2$. In this case, the distance between $\bm{x}_t$ and $\bm{x}_{t+1}$ (the ``step length'') follows a Rayleigh distribution with scale parameter $\sqrt{2} \sigma$. The parameter $\sigma$ of this model can thus be linked to the speed of movement of the animal. It also determines the extent of the region over which the animal can perceive its habitat.

\subsection{Availability radius model}
\label{sec:model4}
The model described by \cite{michelot2018linking} is another special case of the local Gibbs model. In their approach, the density $\phi$ is uniform over a disc of radius $r$ centred on the origin. At each iteration, the point $\bm\mu$ is sampled from a uniform distribution over $\mathcal{D}_r(\bm{x}_t)$, where $\mathcal{D}_r(\bm{x})$ denotes the disc of radius $r$ and centre $\bm{x}$. Then, the endpoint $\bm{x}_{t+1}$ is sampled from $\pi$ truncated to $\mathcal{D}_r(\bm\mu)$. We will refer to $r$ as the ``availability radius'', drawing a parallel with the availability radius model of \cite{rhodes2005spatially}.

That model can be extended to the case where $r$ is chosen at random at each time step from some probability distribution, i.e.\ $r_t \sim p(r \vert \bm\omega)$. The radius $r_t$ characterizes the size of the relocation region at time $t$. In the limit, if $r_t$ tends to infinity, the points become independent samples from the target distribution. This is not a special case of the sampler described in Section \ref{sec:model2}, but the convergence of the sampled locations to the target distribution is preserved. Indeed, each choice of $r$ can be seen as a different sampler, which itself satisfies the detailed balance condition. In Sections \ref{sec:sim3} and \ref{sec:appli2}, we explore the case where $r_t$ arises from a gamma distribution, with shape $\alpha$ and rate $\rho$, i.e.\ $\bm\omega = (\alpha,\rho)$.

The radius $r_t$ determines the spatial scale of the animal's movement and perception: we assume that the animal perceives the whole disc $\mathcal{D}_{r_t}(\bm\mu)$, and can move to any of its points between $t$ and $t+1$. As before, the intermediate point $\bm\mu$ is of no particular interest, but it must be generated uniformly from $\mathcal{D}_{r_t}(\bm{x}_t)$ to obtain a valid MCMC sampler. If we integrate $\bm\mu$ out of the selection of $\bm{x}_{t+1}$, to obtain $p(\bm{x}_{t+1} \vert \bm{x}_t) = \int_\Omega p(\bm{x}_{t+1} \vert \bm\mu) p(\bm\mu \vert \bm{x}_t) d\bm\mu$, we find that the relocation region of the animal at time $t$ (i.e.\ the region over which $p(\bm{x}_{t+1} \vert \bm{x}_t)>0$) is a disc of radius $2r_t$ centred on $\bm{x}_t$.

\begin{figure}[htbp]
	\centering
    \includegraphics[width=0.32\textwidth]{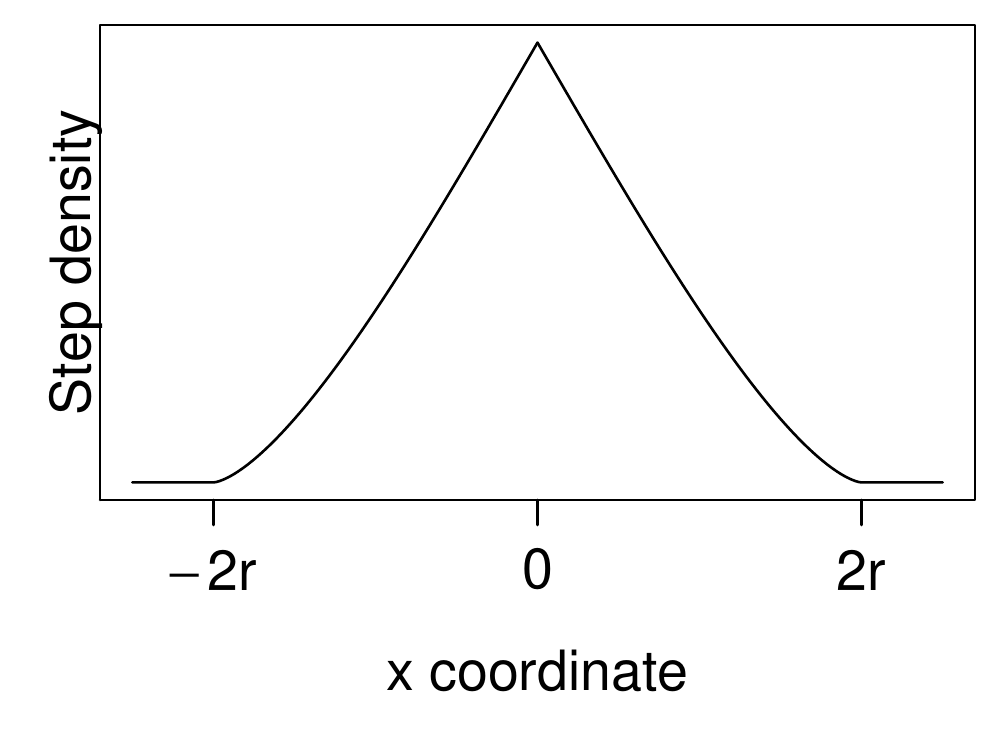}
    \includegraphics[width=0.32\textwidth]{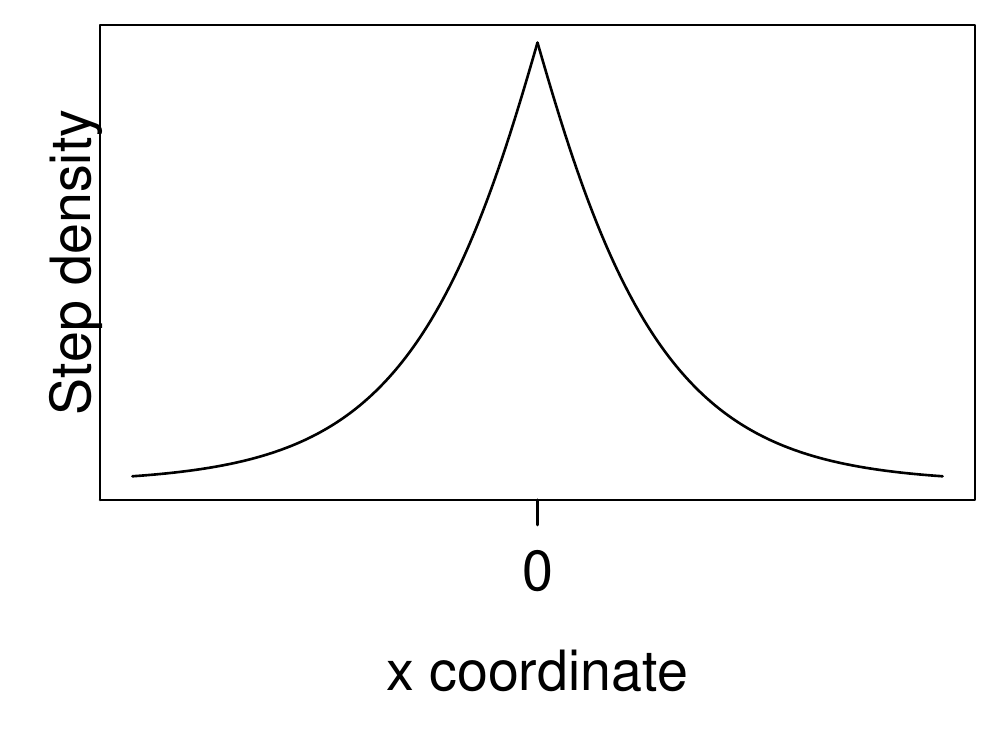}
    \includegraphics[width=0.32\textwidth]{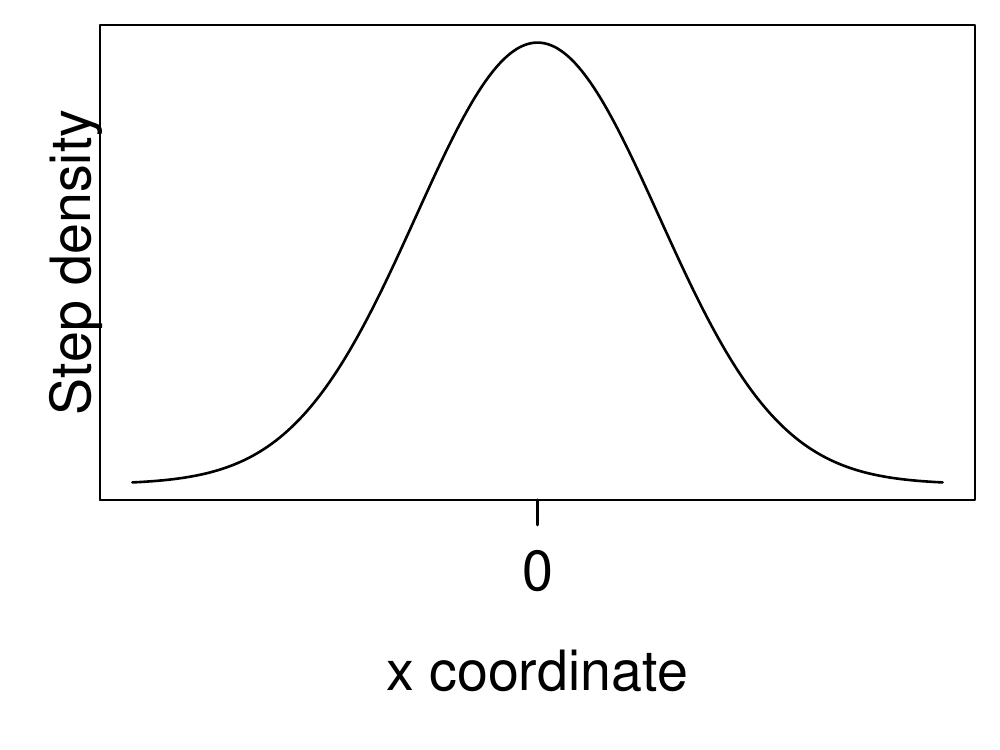}
    \caption{Slices through the two-dimensional resource-independent step densities in three different MCMC movement models. (The step densities are symmetric around the origin.) Left: local Gibbs model with constant availability radius $r$. Middle: local Gibbs model, with availability radius $r_t$ drawn from a gamma distribution. Right: local Gibbs model with normal transition density. Analytical formulas can be obtained for the step densities of the local Gibbs models with constant availability radius or normal transition density, but the middle plot is obtained numerically.}
    \label{fig:stepdensLG}
\end{figure}

Figure \ref{fig:stepdensLG} compares the shapes of the step densities of three local Gibbs model formulations ($r_t$ constant, $r_t \sim$ gamma, and normal kernel), in the absence of covariate effects. This illustrates the flexibility of the underlying movement model.

\subsection{Relationship with Gibbs sampling}

As the name suggests, the sampler defined here has close links with the usual Gibbs sampler. The Gibbs sampler for a target distribution $\pi$ defined on $\Omega$ involves writing $\bm{x} \in \Omega$ as $(x_1, \ldots, x_k)$, and then sampling each new $x_j$ from its `full conditional' distribution $\pi(x_j|x_{(j)})$, that is $\pi$ restricted to the subspace of $\Omega$ defined by fixing $x_{(j)}$. Within that lower-dimensional subspace, the sampling is `global', over the whole range of $x_j$. 
In contrast, the local Gibbs sampler moves within $\Omega$ itself rather than a lower-dimensional subspace, but only `locally' i.e. it only samples $\bm{x}_{t+1}$ from points close to the current point $\bm{x}_{t}$. It is rejection-free because, like the conventional Gibbs sampler, it incorporates the target distribution $\pi$ directly into its sampling distribution. In the simplest case, the `availability radius' model of \cite{michelot2018linking}, the sampling distribution \emph{is} simply the target distribution restricted to a bounded region containing the current point. The region has to be chosen with care, to ensure detailed balance, as described in Section \ref{sec:model4}. Specifically, just sampling $\bm{x}_{t+1}$ from $\pi$ restricted to, say, a symmetric region around $\bm{x}_{t}$ does \emph{not} define a valid sampler from $\pi$. The more general form defined in Section \ref{sec:model2} uses the same concept, but with extra flexibility that enables more realistic matching of the movement of $\bm{x}_{t}$ around $\Omega$ to the animal movement in the application. 

\section{Inference}
\label{sec:estimation}
The parameters of an MCMC step selection model (as described in Section \ref{sec:model}) can be estimated, to carry out inference about the effects of environmental covariates on the animal's distribution in space. We denote by $\bm{\beta} = (\beta_1,\dots,\beta_n)'$ the vector of parameters of the RSF, and $\bm{\theta}$ the vector of movement parameters (i.e.\ tuning parameters of the MCMC sampler). We consider $T$ observed locations $(\bm{x}_1,\dots,\bm{x}_T)$. The likelihood of each step from $\bm{x}_t$ to $\bm{x}_{t+1}$ is given by the transition kernel $p(\bm{x}_{t+1} \vert \bm{x}_t, \bm{\beta}, \bm{\theta})$ of the underlying MCMC algorithm. By the Markov property, the steps are independent, and the complete likelihood is the product of the likelihood of the steps,
\begin{equation} \label{eqn:MCMClike}
	L(\bm{x}_1,\dots,\bm{x}_T \vert \bm{\beta}, \bm{\theta}) = \prod_{t=1}^{T-1} p(\bm{x}_{t+1} \vert \bm{x}_t, \bm{\beta}, \bm{\theta}).
\end{equation}

\subsection{The local Gibbs likelihood}
\label{sec:estimation1}
Our aim is to estimate the habitat selection parameters $\bm{\beta}$ (parameters of the utilisation distribution $\pi$), and the movement parameters $\bm{\theta}$ (parameters of the transition density $\phi$), from observed animal telemetry data. Under the local Gibbs movement model, the likelihood of a step from $\bm{x}$ to $\bm{y}$ can be written as a function of $\bm{\theta}$ and $\bm{\beta}$. It is given by the transition kernel of the algorithm (Equation \ref{eqn:stepdens}) and, for $T$ observed locations $(\bm{x}_1,\dots,\bm{x}_T)$, the full likelihood is obtained as the product of the likelihood for each observed step (Equation \ref{eqn:MCMClike}). Estimates of the model parameters can be obtained by maximizing the likelihood with respect to $\bm{\beta}$ and $\bm{\theta}$. Alternatively, in a Bayesian framework, one could add a penalty term to the likelihood, corresponding to the prior, and maximise the resulting posterior distribution of the parameters (or sample from it).

In this framework, it is straightforward to account for missing location data in the likelihood: missing steps (i.e.\ with missing start point or end point) have no contribution. If several independent movement tracks are collected, their joint likelihood may be calculated as the product of the likelihoods of the individual tracks, to obtain pooled parameter estimates.

\paragraph{Normal kernel model}
The likelihood of a step from $\bm{x}$ to $\bm{y}$, under the normal kernel model, is
\begin{equation*}
	p(\bm{y} \vert \bm{x}, \bm\beta, \sigma) = \pi(\bm{y} \vert \bm\beta) \int_{\bm\mu \in \Omega}
    \dfrac{\varphi(\bm{y} \vert \bm\mu, \sigma^2 \bm{I}_2) \varphi(\bm\mu \vert \bm{x}, \sigma^2 \bm{I}_2)}
        {\int_{\bm{z} \in \Omega} \pi(\bm{z} \vert \bm\beta) \varphi(\bm{z} \vert \bm\mu, \sigma^2 \bm{I}_2) d\bm{z}}
        d\bm\mu,
\end{equation*}
where $\varphi(\cdot \vert \bm\mu, \bm\Sigma)$ is the multivariate normal density with mean $\bm\mu$ and covariance matrix $\bm\Sigma$, and $\bm{I}_2$ is the $2 \times 2$ identity matrix.

\paragraph{Random availability radius model}
We consider the availability radius model presented in Section \ref{sec:model4}, in the case where the radius $r$ is random, with distribution $p(r \vert \bm\omega)$. The likelihood of this model can be written in terms of the parameters $\bm\omega$ of the distribution of the radius $r$. The transition density is uniform on a disc centred on the origin, with radius parameter $r$. That is, for any points $\bm{a}$ and $\bm{b}$, and using the notation of Section \ref{sec:model4},
\begin{equation*}
	\phi(\bm{b} \vert \bm{a}, r) = \dfrac{1}{\pi r^2} I_{\{\bm{b} \in D_r(\bm{a})\}},
\end{equation*}
where $I_A$ is the indicator function for event $A$. Following from Equation \ref{eqn:stepdens2}, the likelihood of a step from $\bm{x}$ to $\bm{y}$ is therefore
\begin{align} \label{eqn:RARlike}
	p(\bm{y} \vert \bm{x}, \bm\beta, \bm\omega) & = \pi(\bm{y} \vert \bm\beta)
    	\int_{r=0}^{\infty} p(r \vert \bm\omega)
        \int_{\bm\mu \in \Omega} 
        \dfrac{1}{\pi r^2} \dfrac{I_{\{ \bm{y} \in D_r(\bm\mu) \}} I_{\{ \bm\mu \in D_r(\bm{x}) \}}}
        {\int_{\bm{z} \in \Omega} \pi(\bm{z} \vert \bm\beta) I_{\{ \bm{z} \in D_r(\bm\mu) \}} d\bm{z}} 
        d\bm\mu dr \nonumber \\
        & = \dfrac{\pi(\bm{y} \vert \bm\beta)}{\pi}
    	\int_{r=d/2}^{\infty} \dfrac{p(r \vert \bm\omega)}{r^2}
        \int_{\bm\mu \in \mathcal{D}_r^{(\bm{x},\bm{y})}} 
        \dfrac{1}{\int_{\bm{z} \in \mathcal{D}_r(\bm\mu)} \pi(\bm{z} \vert \bm\beta) d\bm{z}} 
        d\bm\mu dr,
\end{align}
where $d = \lVert \bm{y} - \bm{x} \rVert$ is the distance between $\bm{x}$ and $\bm{y}$, and $\mathcal{D}_r^{(\bm{x},\bm{y})} = \mathcal{D}_r(\bm{x}) \cap \mathcal{D}_r(\bm{y})$. The integral over $r$ is between $d/2$ and $\infty$ because the step length $d$ cannot be longer than $2r$.

\subsection{Monte Carlo approximation of the likelihood}
\label{sec:estimation2}
The integrals of Equation \ref{eqn:stepdens} cannot generally be evaluated analytically. Monte Carlo sampling can be used as follows to approximate the likelihood of a step from $\bm{x}$ to $\bm{y}$. 

For $i \in \{ 1, \dots, n_c \}$, sample $\bm\mu_i$ from $\phi(\cdot \vert \bm{x}, \bm\theta)$ and, for $j \in \{ 1, \dots, n_z \}$, sample $\bm{z}_{ij}$ from $\phi(\cdot \vert \bm\mu_i, \bm\theta)$. Then, the likelihood of a step from $\bm{x}$ to $\bm{y}$, given in Equation \ref{eqn:stepdens}, can be approximated by
\begin{equation} \label{eqn:MClike}
	\hat{p}(\bm{y} \vert \bm{x}, \bm\beta, \bm\theta) = \pi(\bm{y} \vert \bm\beta) \dfrac{n_z}{n_c}
    	\sum_{i=1}^{n_c} \dfrac{\phi(\bm{y} \vert \bm\mu_i, \bm\theta)}
        {\sum_{j=1}^{n_z} \pi(\bm{z}_{ij} \vert \bm\beta)}.
\end{equation}

In the case where $\bm\theta$ is random, the likelihood is written with one additional integral (Equation \ref{eqn:stepdens2}), which must also be approximated. As an example, the approximate likelihood of the random availability radius model is given in Appendix 1.

The approximation in Equation \ref{eqn:MClike} can be made arbitrarily accurate by choosing large sizes of Monte Carlo samples ($n_c$ and $n_z$). Latin hypercube sampling can be used to reduce the number of samples needed in the approximation of the likelihood \citep{mckay1979comparison}. In Appendix 1, we describe the practical implementation of the local Gibbs approximate likelihood for the normal kernel model and the random availability radius model.

Similar Monte Carlo approximations can be found in most SSF analyses, when conditional logistic regression is used to compare each observed step of the animal to \emph{potential} steps that the animal could have selected, starting from the same point. Comparing actual endpoints with potential endpoints in terms of their covariate values allows inference on habitat selection by the animal. In practice, for each observed step $(\bm{x}_t,\bm{x}_{t+1})$, a set of random steps with origin $\bm{x}_t$ is sampled from the underlying movement model. This technique is used to approximate the integral of the SSF over the movement kernel, i.e.\ the denominator of the SSF likelihood (Equation \ref{eqn:ssf}). The number $K$ of random steps varies a lot between studies, between $K=2$ and $K=200$, according to \cite{thurfjell2014applications}. Note that we are likely to need much larger Monte Carlo samples for the approximation of the local Gibbs likelihood, because of the nested integrals (Equation \ref{eqn:stepdens}).

\subsection{State-switching model}
\label{sec:MSLG}
The local Gibbs model is formulated in terms of a set of movement parameters $\bm\theta$, the parameters of the transition density $\phi$ in Section \ref{sec:model2}. For example, the scale of the animal's movement is characterised by the variance $\sigma^2$ in the normal kernel case, and the availability radius $r$ (or the parameters of its distribution if $r$ is random) in the availability radius model. This framework can be extended by considering that the animal can switch between $N$ discrete states through time, each associated with a set of movement parameters ($\bm\theta^{(1)},\dots,\bm\theta^{(N)}$). We introduce an latent state process $S_t \in \{ 1, \dots, N \}$, which defines which state is active at each time step $t$. Multistate models have become very popular in movement ecology, to describe animal movement as the consequence of behaviour. The states are usually treated as proxies for behavioural states of the animal, such as ``foraging'' or ``exploring'' \citep{morales2004extracting, langrock2012flexible}.

The target distribution of the local Gibbs sampler does not depend on the movement parameters $\bm\theta$. In this multistate formulation, the movement process switches between $N$ local Gibbs models, all with the same utilisation distribution $\pi$. The overall  utilisation distribution of the state-switching model is therefore also $\pi$. The underlying MCMC algorithm can be seen as a hybrid algorithm, based on $N$ transition kernels which share the same stationary distribution \citep{tierney1994markov}.

If $S_t$ is chosen to be a Markov chain, this defines a hidden Markov model, and the associated inferential machinery can be used \citep{zucchini2016hidden}. In particular, the likelihood of the state-switching local Gibbs model can be calculated with the forward algorithm, which provides an efficient way to sum over all possible state sequences. In the present context, it can be written in the form of the following matrix product,
\begin{equation*}
	L(\bm{x}_1, \dots, \bm{x}_T \vert \bm\beta, \{ \bm\theta^{(j)} \} ) = \bm\delta^{(0)} 
    	\bm{P}(\bm{x}_1,\bm{x}_2) \bm\Gamma \bm{P}(\bm{x}_2,\bm{x}_3) \cdots
        \bm\Gamma \bm{P}(\bm{x}_{T-1},\bm{x}_T) \bm{1}',
\end{equation*}
where $\bm\delta^{(0)}$ is the initial distribution of the Markov chain, $\bm\Gamma = (\gamma_{ij})_{i,j=1}^N$ is its transition probability matrix, $\bm{P}(\bm{x}_t,\bm{x}_{t+1})$ is the $N \times N$ diagonal matrix with elements $\{ p(\bm{x}_{t+1} \vert \bm{x}_t, \bm\beta, \bm\theta^{(j)}) \}_{j=1}^N$, and $\bm{1}$ is a $N$-vector of ones. The approximate likelihood of a step, given in Equation \ref{eqn:MClike}, can then be plugged into $\bm{P}(\bm{x}_t,\bm{x}_{t+1})$.

Maximum likelihood estimation can be used to obtain estimates of all model parameters, including habitat selection parameters ($\bm\beta$), movement parameters ($\bm\theta^{(j)}$), and transition probabilities ($\bm\Gamma$). The Viterbi algorithm can be implemented to derive the most likely sequence of underlying states, given the data and a fitted model \citep{zucchini2016hidden}. It is used in animal movement analyses to classify observed locations into behavioural phases, described by different movement characteristics \citep{michelot2016movehmm}.

\cite{roever2014pitfalls} showed that ignoring animal behaviour in habitat selection studies could lead to incorrect conclusions. They argued for a two-stage modelling approach, in which tracks would first be classified into behavioural states using a state-switching correlated random walk model \citep{morales2004extracting}, and a separate set of habitat selection parameters would then be estimated for each behavioural state. The state-switching local Gibbs model that we suggest here is different, because it estimates only one set of habitat selection parameters for all states. However, the scale of perception and movement can differ among the states, if they are characterised by different parameters $\bm\theta^{(j)}$. Then, the state-switching local Gibbs model does account for the behavioural heterogeneity in the scale of habitat selection.

\section{Simulation study}
\label{sec:sim}
In this section, we investigate the performance of the method to estimate the RSF and the movement parameters from simulated movement data. These simulations are designed to resemble the data used in the case study of Section \ref{sec:appli}. The R code used for the simulations is available in the supplementary material.

\begin{figure}[htbp]
	\centering
    \includegraphics[width=0.51\textwidth]{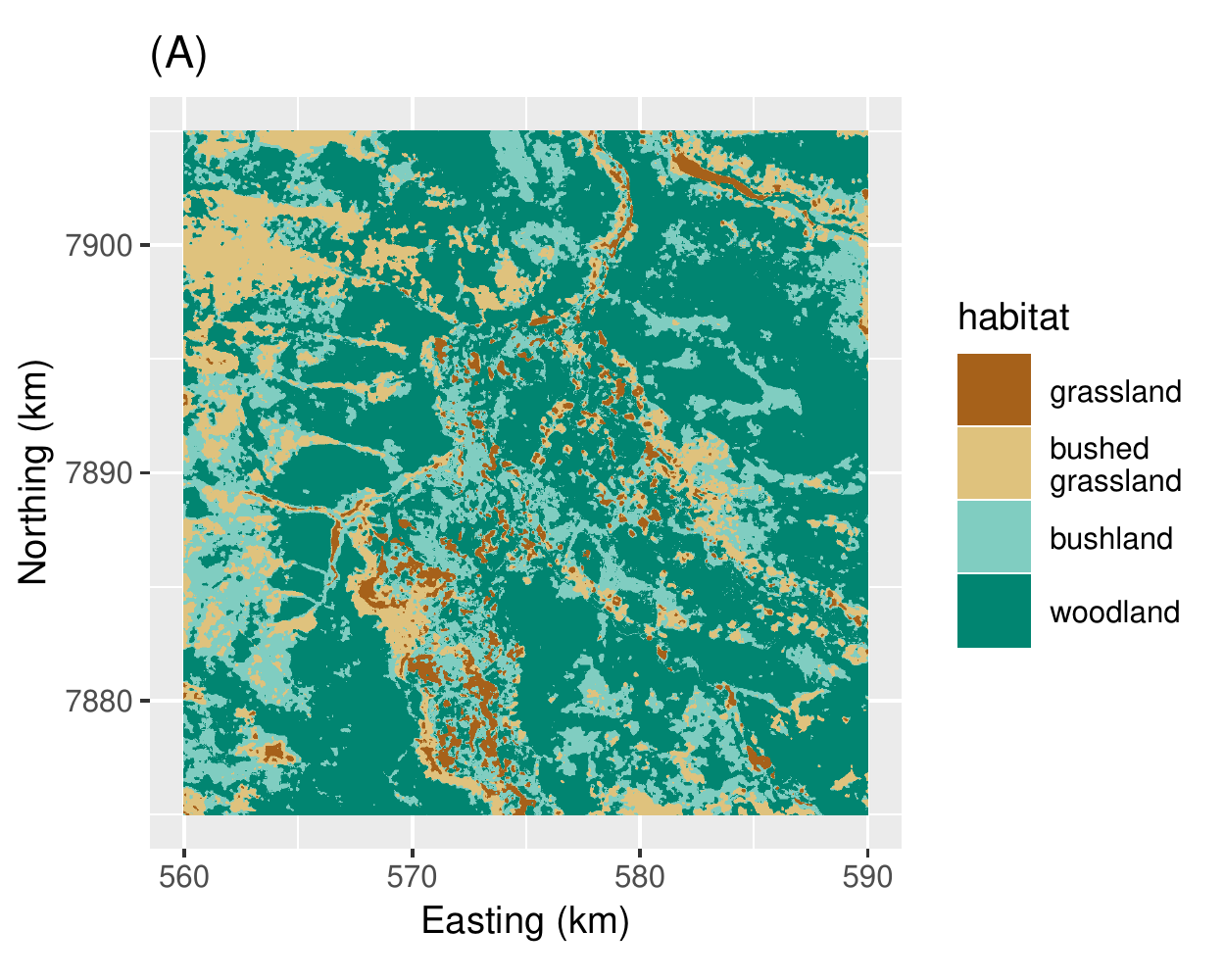}
    \includegraphics[width=0.48\textwidth]{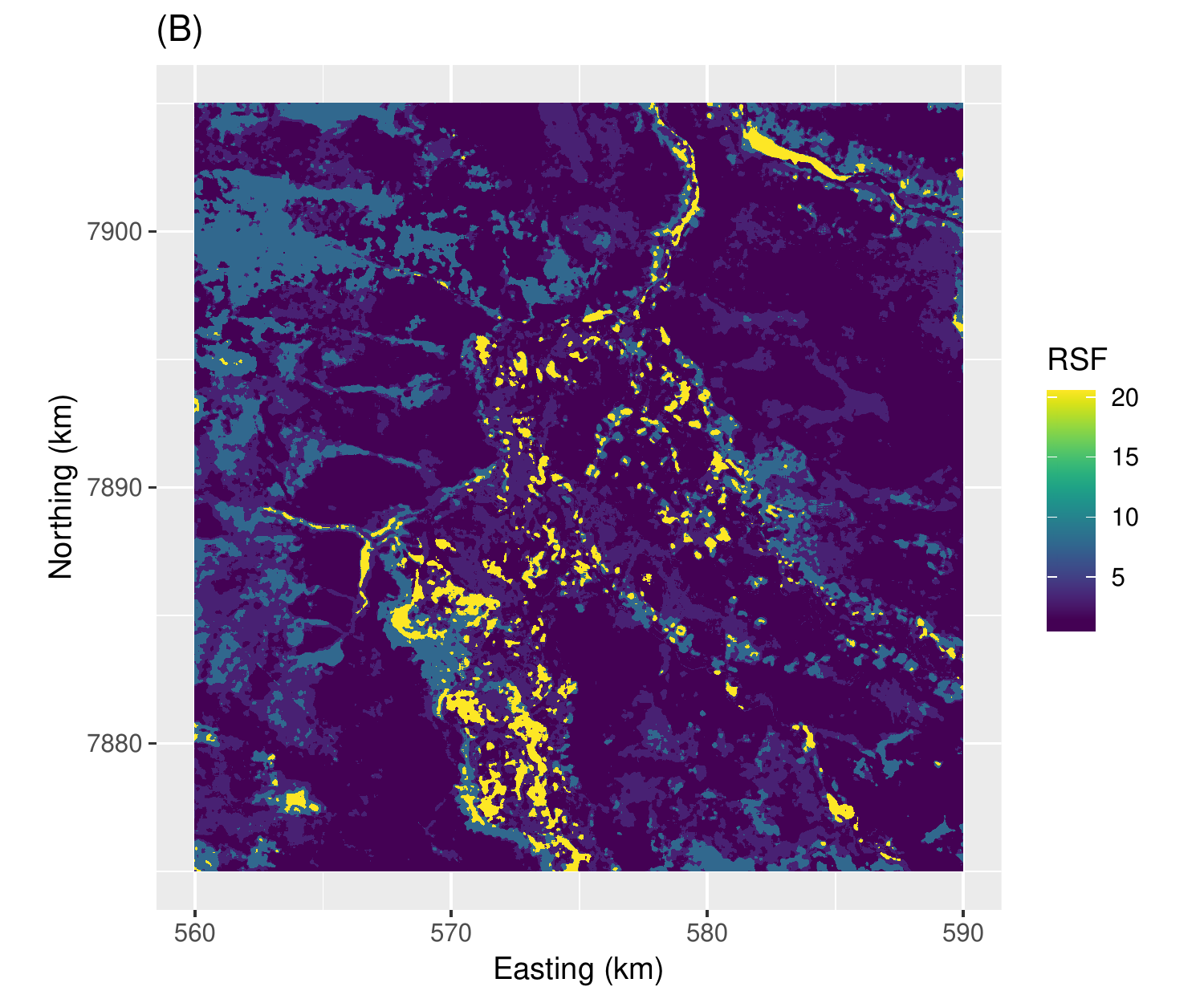}
    \caption{(A) Map of habitats. (B) Artificial resource selection function used in the simulation studies, constructed from the habitat map with selection parameters $\bm{\beta} = (3,2,1,0)'$. Colour figures can be viewed in the online version of the paper.}
    \label{fig:simrsf}
\end{figure}

We used a categorical vegetation map from Hwange National Park (Zimbabwe) as baseline environmental layer for our simulations. The map covers 30km$\times$30km at a 30m resolution. The vegetation type takes four values: grassland, bushed grassland, bushland, and woodland. For all simulations, we considered the artificial RSF defined as $w(\bm{c}) = \exp(\bm{\beta}'\bm{c})$, with $\bm{\beta} = (\beta_\text{G}, \beta_\text{BG}, \beta_\text{B}, \beta_\text{W})' = (3,2,1,0)'$. Because the covariate is categorical, the effect of each habitat type is modelled relatively to the effect of the reference category, here chosen to be the woodland habitat. Maps of the habitats and of the RSF are shown in Figure \ref{fig:simrsf}.

We considered three scenarios, corresponding to three different movement models: (1) the local Gibbs model with normal transition density, (2) the state-switching local Gibbs model with normal transition density, and (3) the local Gibbs model with random availability radius. In each scenario, we simulated 50 movement tracks from the local Gibbs algorithm with target distribution the normalized RSF (Equation \ref{eqn:ud}). Each trajectory was of length $T=1000$ observations, and the initial location was sampled from the target distribution. We rejected any trajectory that did not visit all four habitat types, to ensure that all habitat selection parameters could be estimated. For each of the 50 simulated tracks, we used the R function \texttt{optim} to maximise the approximate likelihood of the model, given in Section \ref{sec:estimation2}, to recover the values of the parameters.

\subsection{Scenario 1: normal kernel model}
\label{sec:sim1}
There were four unknown parameters to estimate: three regression coefficients for the RSF (the woodland habitat was chosen as the reference category, with a coefficient set to zero), and the variance parameter of the normal transition density. We chose $n_c=n_z=50$ as the sizes of the Monte Carlo samples in the approximation of the likelihood. This choice was based on preliminary experiments with various sizes of Monte Carlo samples; 50 was a good trade-off between accuracy and computational speed. We used Latin hypercube sampling to obtain random uniform Monte Carlo samples, and transformed them to normal samples using the quantile function of the normal distribution.

\begin{figure}[htbp]
	\centering
    \includegraphics[width=0.52\textwidth]{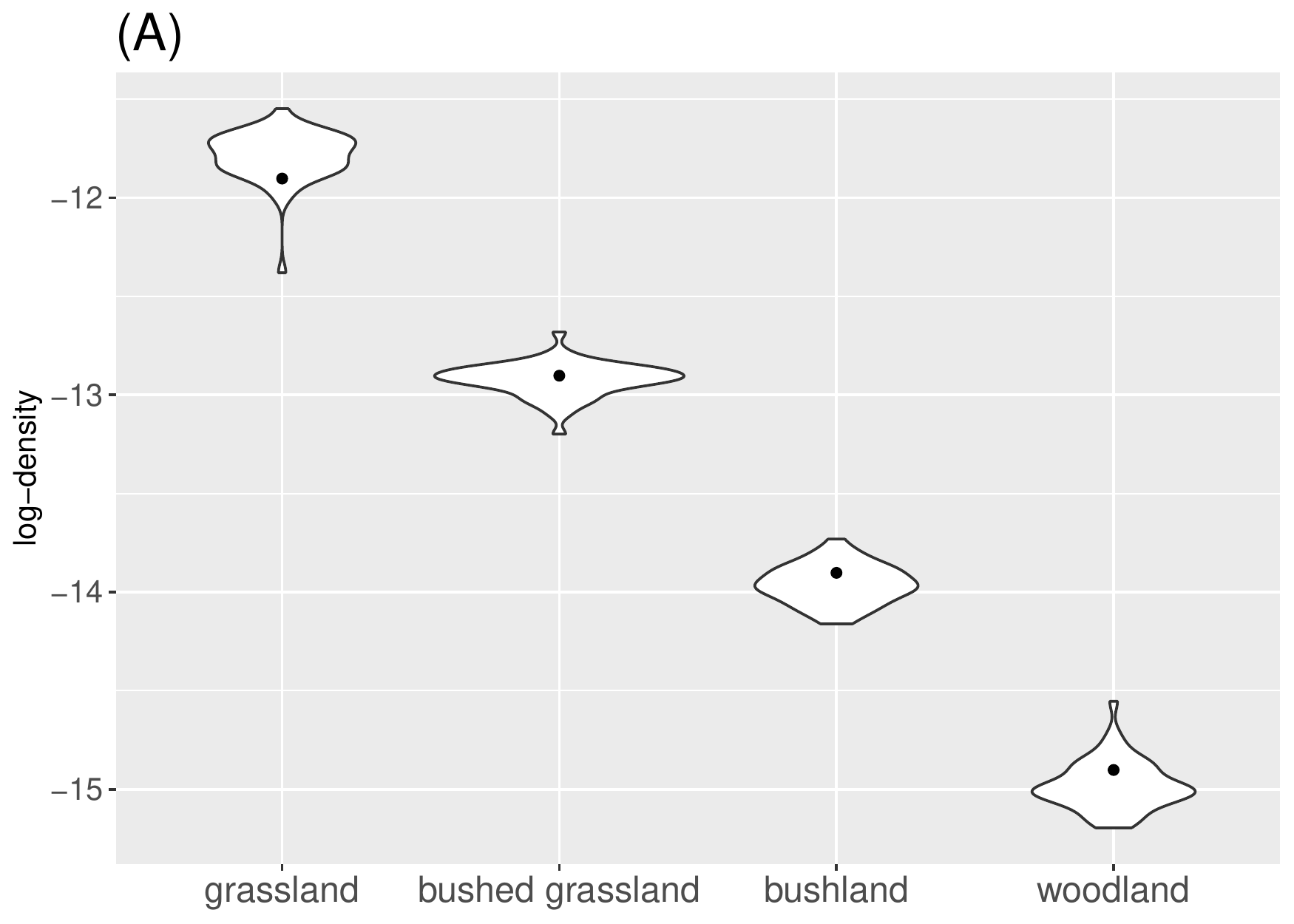}
    \includegraphics[width=0.46\textwidth]{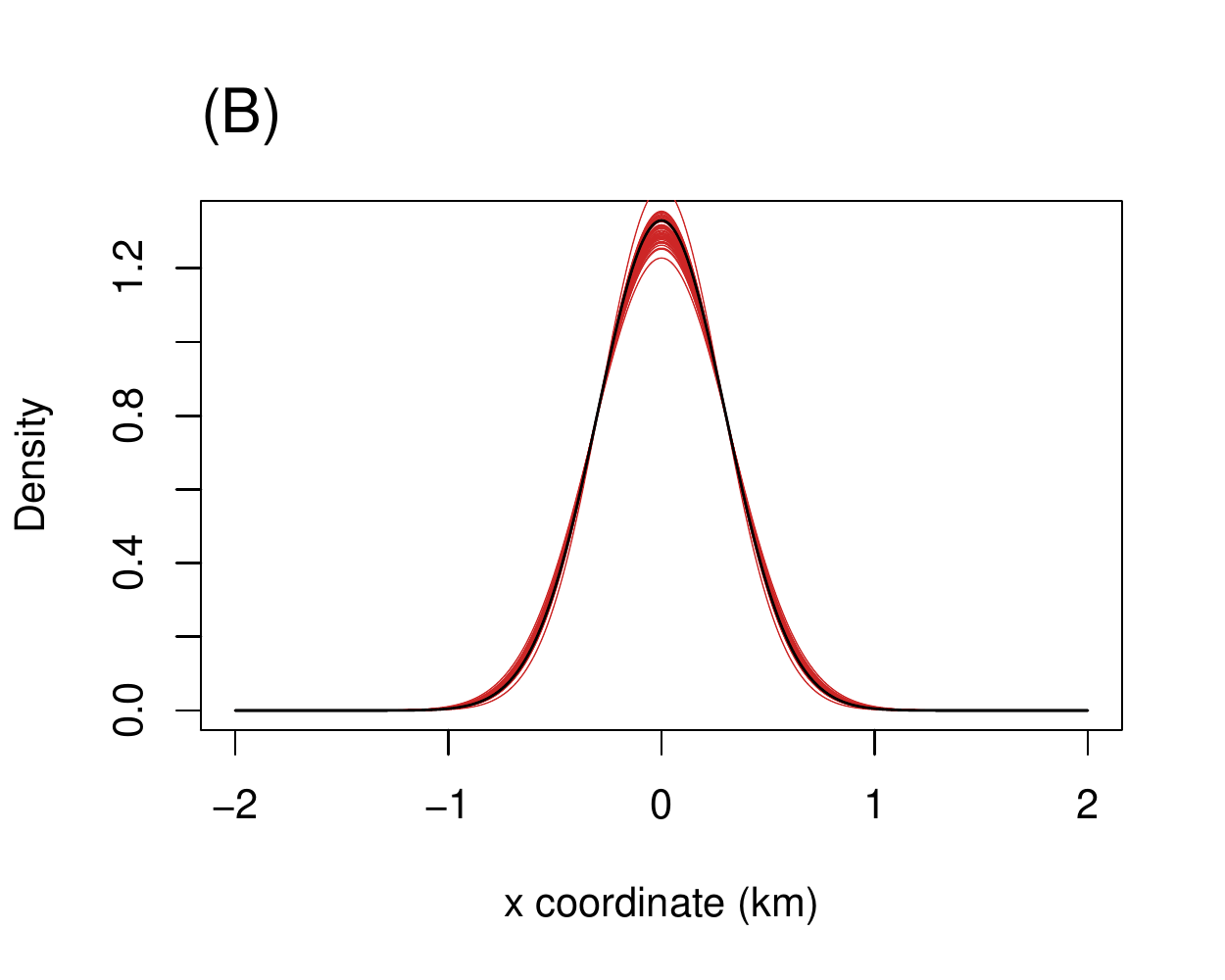}
    \caption{Results of the 50 simulations, for the local Gibbs model with normal transition density. (A) The violin plots show the 50 values of the estimated densities for each of the four habitats (as defined in Equation \ref{eqn:estUD}; here on the log-scale). The black dots indicate the true values. The y-axis is on a log-scale, so that all the densities can be visualised on the same plot. (B) The red lines are the 50 estimated normal transition densities, and the black line is the true density used in the simulations. Colour figures can be viewed in the online version of the paper.}
    \label{fig:simres1}
\end{figure}

With categorical environmental data, the utilisation distribution is constant over each category (here, each habitat type), so we can derive the true and estimated utilisation value of each habitat,
\begin{equation}
	\pi_i = \dfrac{\exp(\beta_i)}{\int_{\bm{x} \in \Omega} \exp(\bm{\beta}' \bm{c}(\bm{x})) d\bm{x}},\quad 	
    \text{and } \hat{\pi}_i = 
    \dfrac{\exp(\hat{\beta}_i)}{\int_{\bm{x} \in \Omega} \exp(\hat{\bm{\beta}}' \bm{c}(\bm{x})) d\bm{x}},
    \label{eqn:estUD}
\end{equation}
where $i \in \{$G,BG,B,W$\}$ is the index of the habitat type, and the $\hat\beta_i$ are the estimated RSF parameters. Figure \ref{fig:simres1}(A) shows a comparison of the estimated and true utilisation values for the four habitat types. By comparing the utilisation values (rather than the true and estimated parameters), we directly assess the resemblance of the true and estimated utilisation distributions. The true shape of the utilisation distribution was adequately captured by the estimated RSFs in most of the simulations. Figure \ref{fig:simres1}(B) shows the 50 estimated transition densities. They all closely resemble the true distribution used for the simulations, indicating that the movement component of the model was successfully captured in the estimation.

\subsection{Scenario 2: state-switching normal kernel model}
\label{sec:sim2}
We ran 50 simulations for the state-switching local Gibbs model with normal transition density described in Section \ref{sec:MSLG}. We used a 2-state model with one movement parameter for each state: $(\sigma_1, \sigma_2) = (0.2,1)$. In this scenario, the two states capture slow and fast movement, respectively. We chose the transition probabilities $\gamma_{11}=\gamma_{22}=0.9$, to imitate the tendency of animals to persist in a behavioural state for several time steps. As in Section \ref{sec:sim1}, we used Monte Carlo samples of size $n_c = n_z = 50$ for the approximation of the likelihood.

\begin{figure}[htbp]
	\centering
    \includegraphics[width=0.52\textwidth]{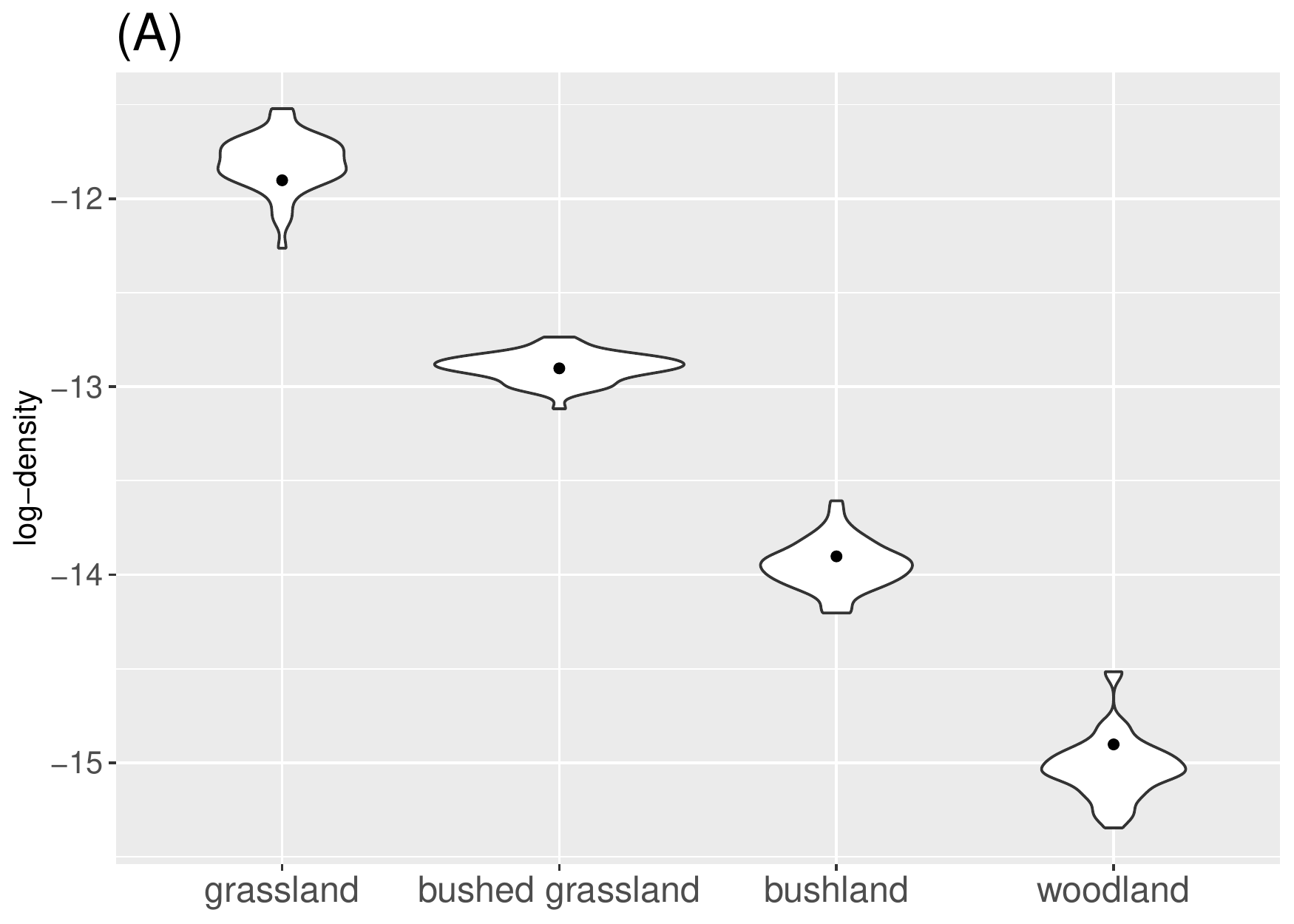}
    \includegraphics[width=0.46\textwidth]{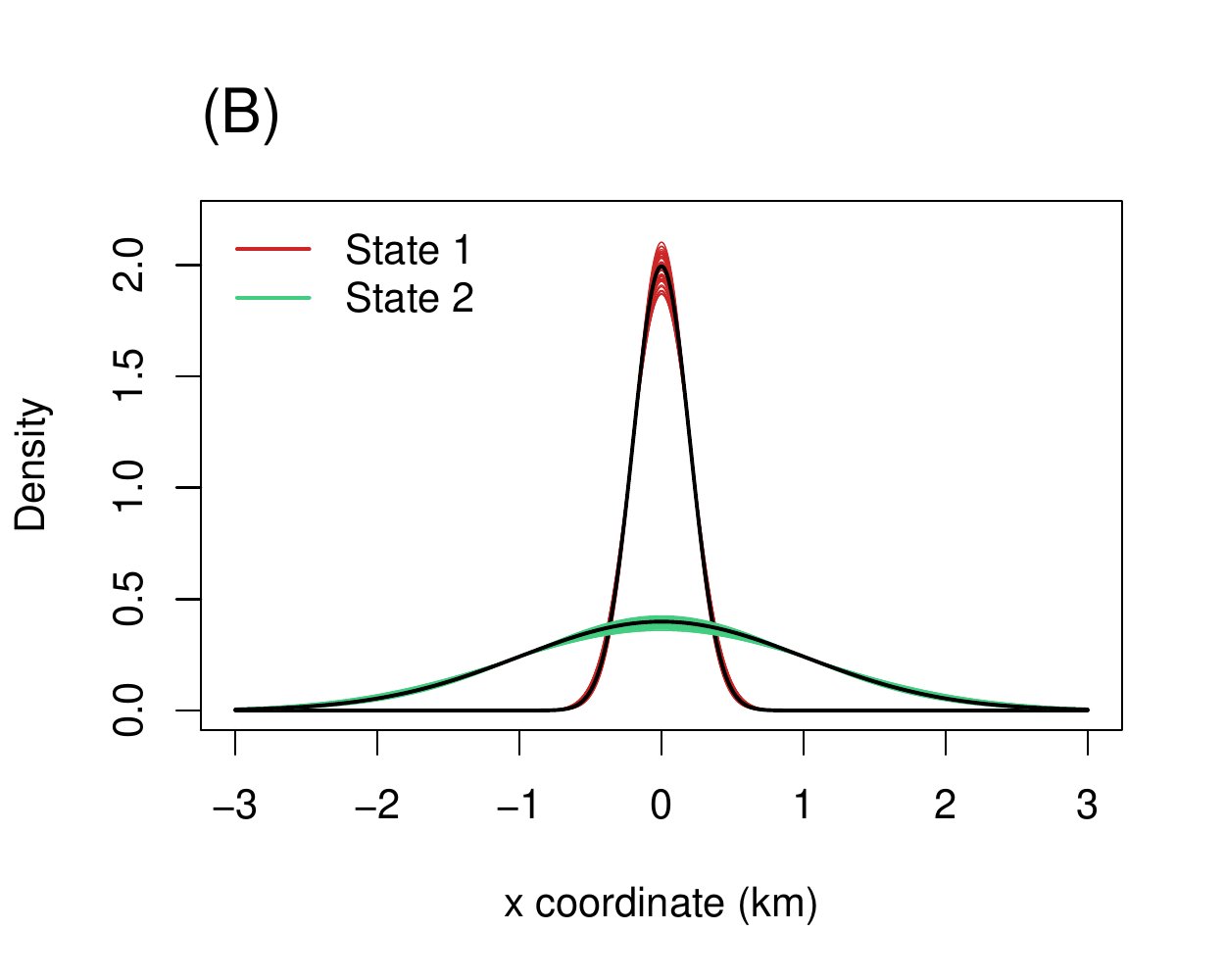}
    \caption{Results of the 50 simulations, for the state-switching normal kernel model. (A) The violin plots show the 50 values of the estimated densities for each of the four habitats (as defined in Equation \ref{eqn:estUD}; here on the log-scale). The black dots indicate the true values. The y-axis is on a log-scale, so that all the densities can be visualised on the same plot. (B) The coloured lines are the 50 estimated transition densities in each state, and the black lines are the true densities used in the simulations. Colour figures can be viewed in the online version of the paper.}
    \label{fig:simres3}
\end{figure}

The estimates of the utilisation values of the four habitats are shown in Figure \ref{fig:simres3}. The utilisation distribution was adequately captured. Figure \ref{fig:simres3} also shows the estimated step densities in the two behavioural states, which all closely match the true step densities used in the simulations. We used the Viterbi algorithm to derive the most likely state sequence for each simulated track, and compared it to the true (simulated) sequence of states. Overall, about 95\% of the steps were correctly classified.

\subsection{Scenario 3: random availability radius model} 
\label{sec:sim3}
We ran a similar simulation study based on the random availability radius model presented in Section \ref{sec:model4}. We generated the availability radius independently at each time step from a gamma distribution with shape parameter $\alpha=0.7$ and rate parameter $\rho=3$. We chose these parameter values to obtain movement speeds that were similar to those observed in the zebra tracking data used in Section \ref{sec:appli}. We simulated 50 tracks of length 1000 from the model, and used maximum likelihood estimation to recover the five parameters ($\beta_\text{G}$, $\beta_\text{BG}$, $\beta_\text{B}$, $\alpha$, and $\rho$). In the Monte Carlo approximation of the likelihood, we used $n_r = n_c = n_z = 30$, based on preliminary experiments with various Monte Carlo sample sizes.

\begin{figure}[htbp]
	\centering
    \includegraphics[width=0.52\textwidth]{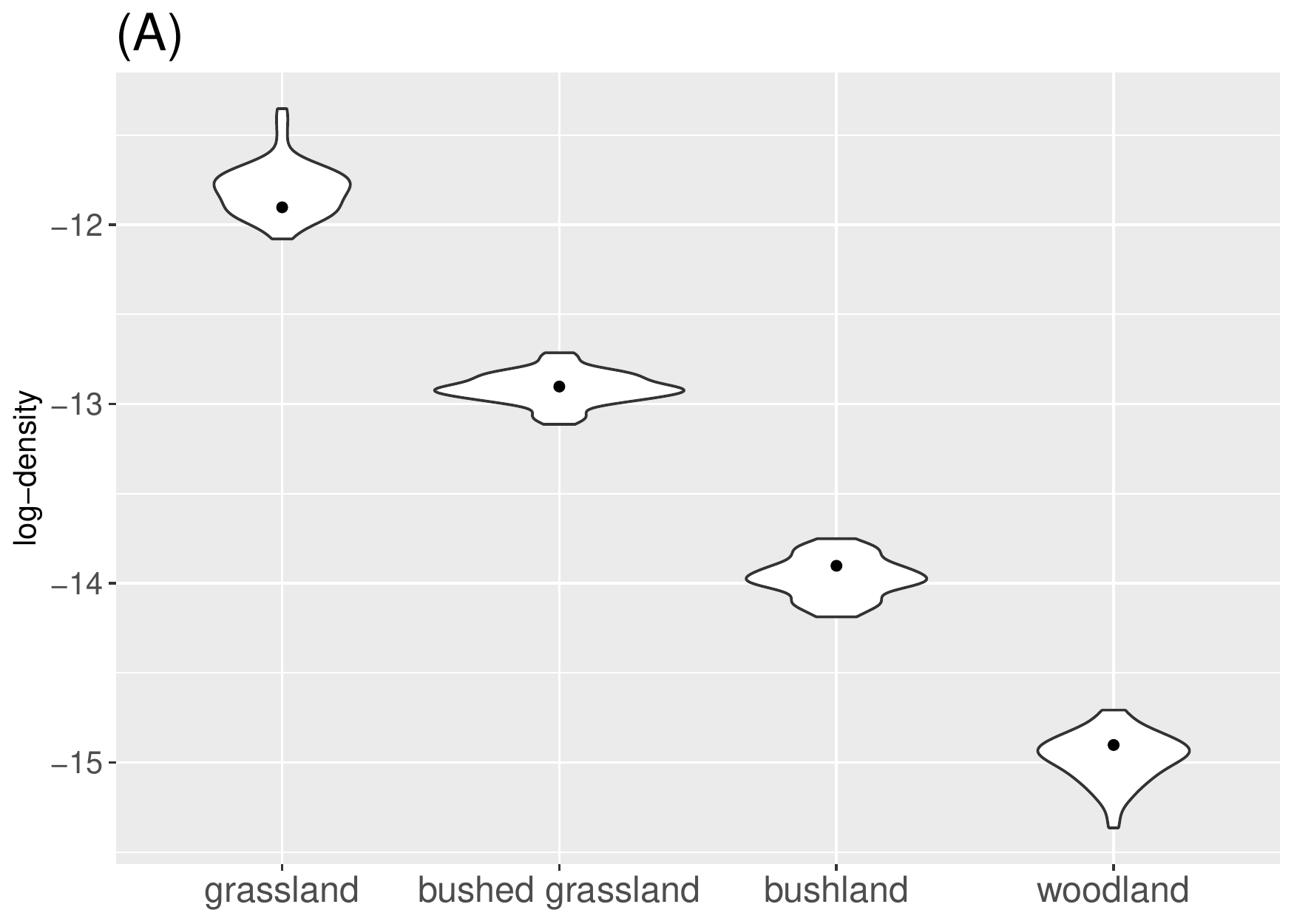}
    \includegraphics[width=0.46\textwidth]{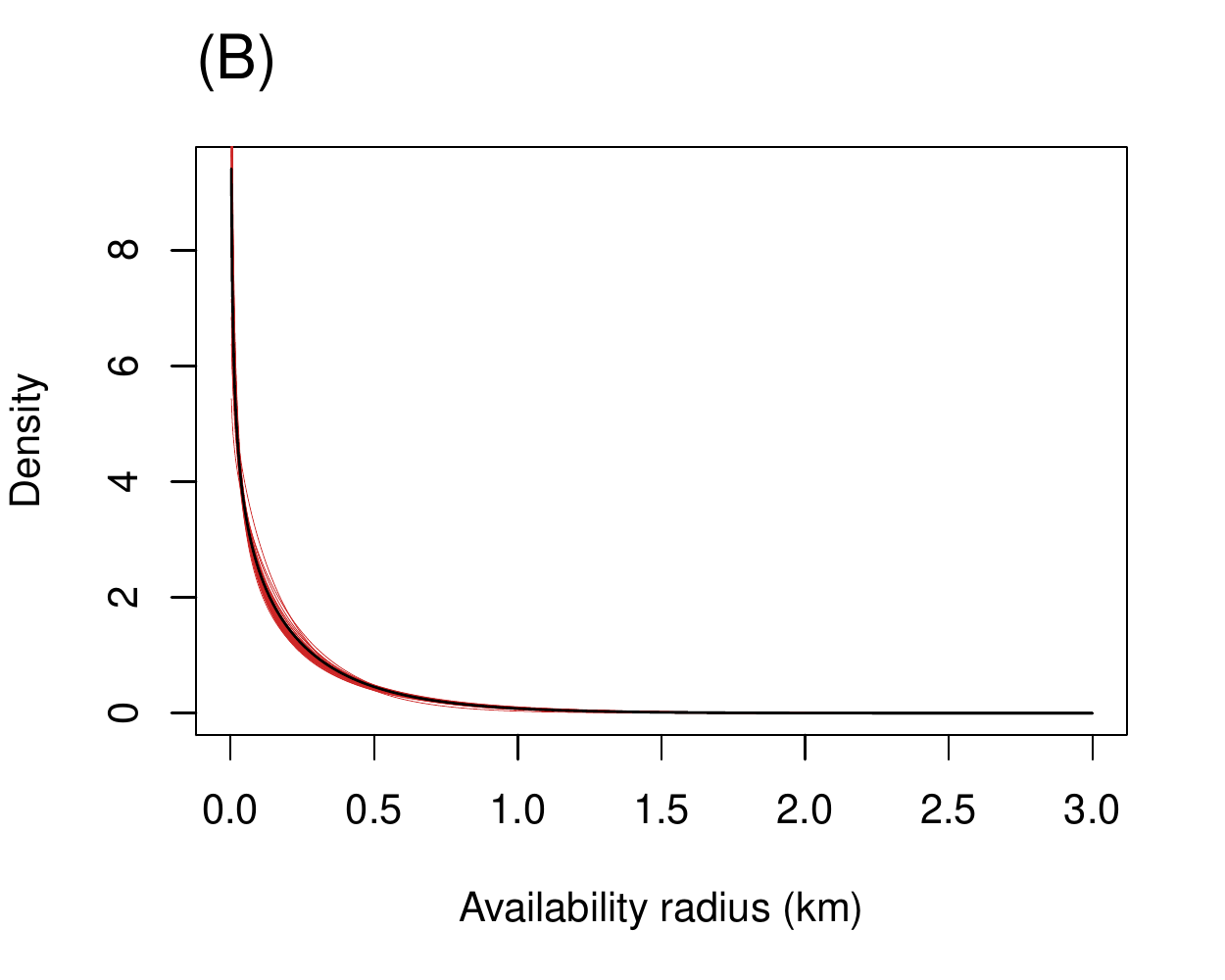}
    \caption{Results of the 50 simulations, for the local Gibbs model with random availability radius. (A) The violin plots show the 50 values of the estimated densities for each of the four habitats (as defined in Equation \ref{eqn:estUD}; here on the log-scale). The black dots indicate the true values. The y-axis is on a log-scale, so that all the densities can be visualised on the same plot. (B) The red lines are the 50 estimated densities of the availability radius, and the black line is the true density used in the simulations. Colour figures can be viewed in the online version of the paper.}
    \label{fig:simres2}
\end{figure}

Figure \ref{fig:simres2} shows the estimated utilisation values, and the estimated availability radius distributions. The estimated utilisation distributions closely capture the true underlying distribution, and the gamma density of the availability radius was successfully estimated in all simulations.

These results confirm that the parameters of the local Gibbs model can be estimated by numerical optimisation of the approximate likelihood.

\section{Application: zebra case study}
\label{sec:appli}
We consider a trajectory of GPS locations of one plains zebra (\emph{Equus quagga}), acquired every 30 minutes from January to May 2014 in Hwange National Park (Zimbabwe). The track consists of 7246 regularly-spaced locations, with 125 missing observations. The habitat layer used to estimate the habitat selection process is a vegetation map, with the same categories as in Section \ref{sec:sim}. A map of the habitat and of the track is shown in Figure \ref{fig:zeb1}(A). The code and data used in the case study are provided in the supplementary material.

\subsection{Normal transition density model}
\label{sec:appli1}
We fitted the local Gibbs model with normal transition density to the track, using the function \texttt{optim} in R to numerically optimise the (approximate) log-likelihood function. We chose $n_c = n_z = 50$ for the Monte Carlo samples in the approximation of the likelihood function (Equation \ref{eqn:MClike}), which was sufficient in the simulation study of Section \ref{sec:sim}.

A numerical optimiser is susceptible to becoming stuck in a local maximum of the likelihood function, and failing to find its global maximum. To circumvent this problem, we fitted the model 50 times, starting from randomly-chosen initial parameter values, and we selected the parameter estimates leading to the best (largest) maximum likelihood. Each model fit took about 8 minutes on a 2GHz i5 CPU. For the best fitting model, we evaluated the Hessian matrix of the log-likelihood function at the maximum likelihood estimate, with the R package numDeriv \citep{gilbert2016numderiv}, and we derived standard errors for the estimated parameters. The approximation of the Hessian also depends on the size of the Monte Carlo samples. We computed it with Monte Carlo samples of increasing sizes, until the estimated standard errors stabilised, around $n_c = n_z = 50$, to ensure that the approximation error was small.

\begin{figure}[hbtp]
	\centering
    \includegraphics[width=0.52\textwidth]{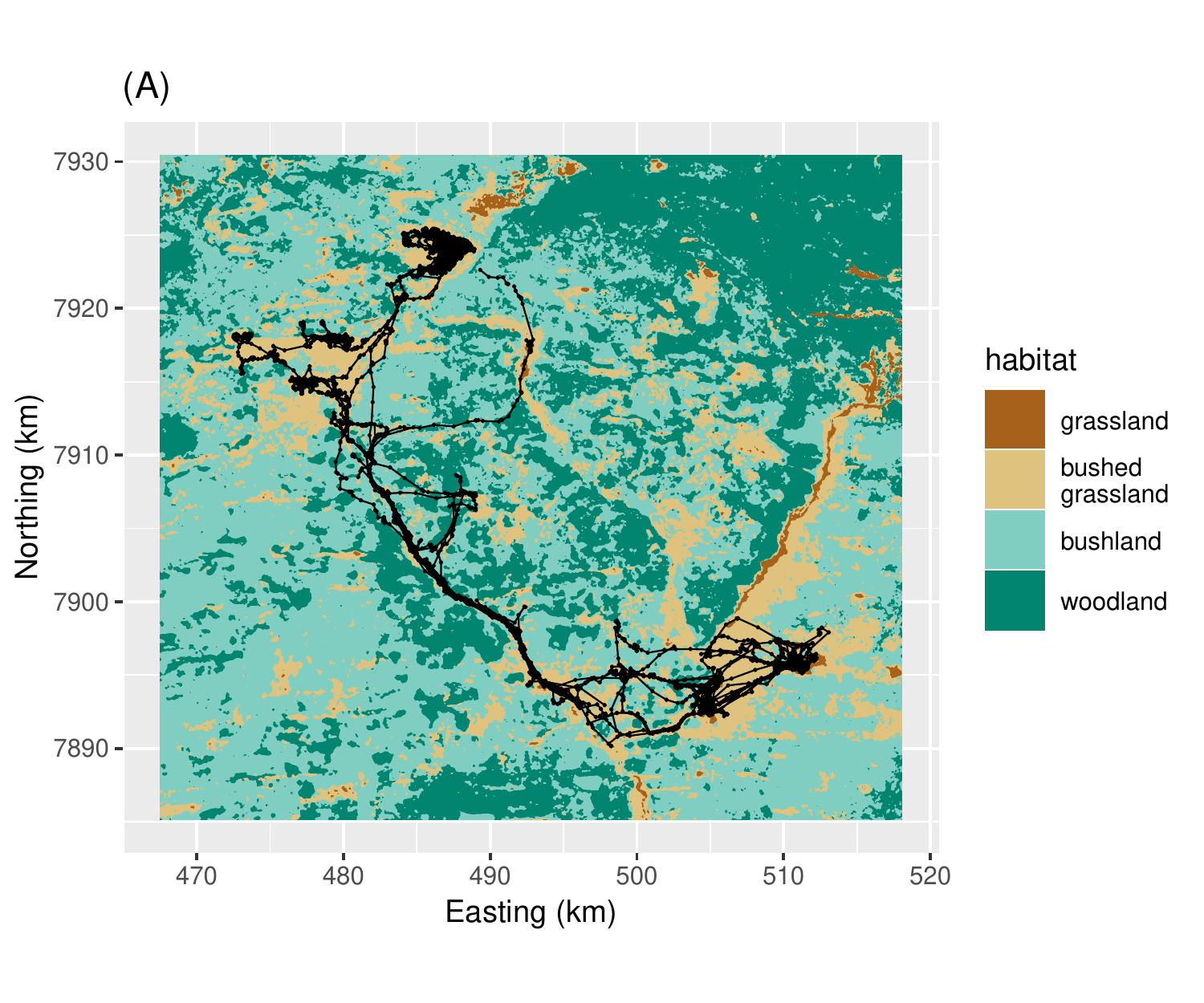}
    \includegraphics[width=0.47\textwidth]{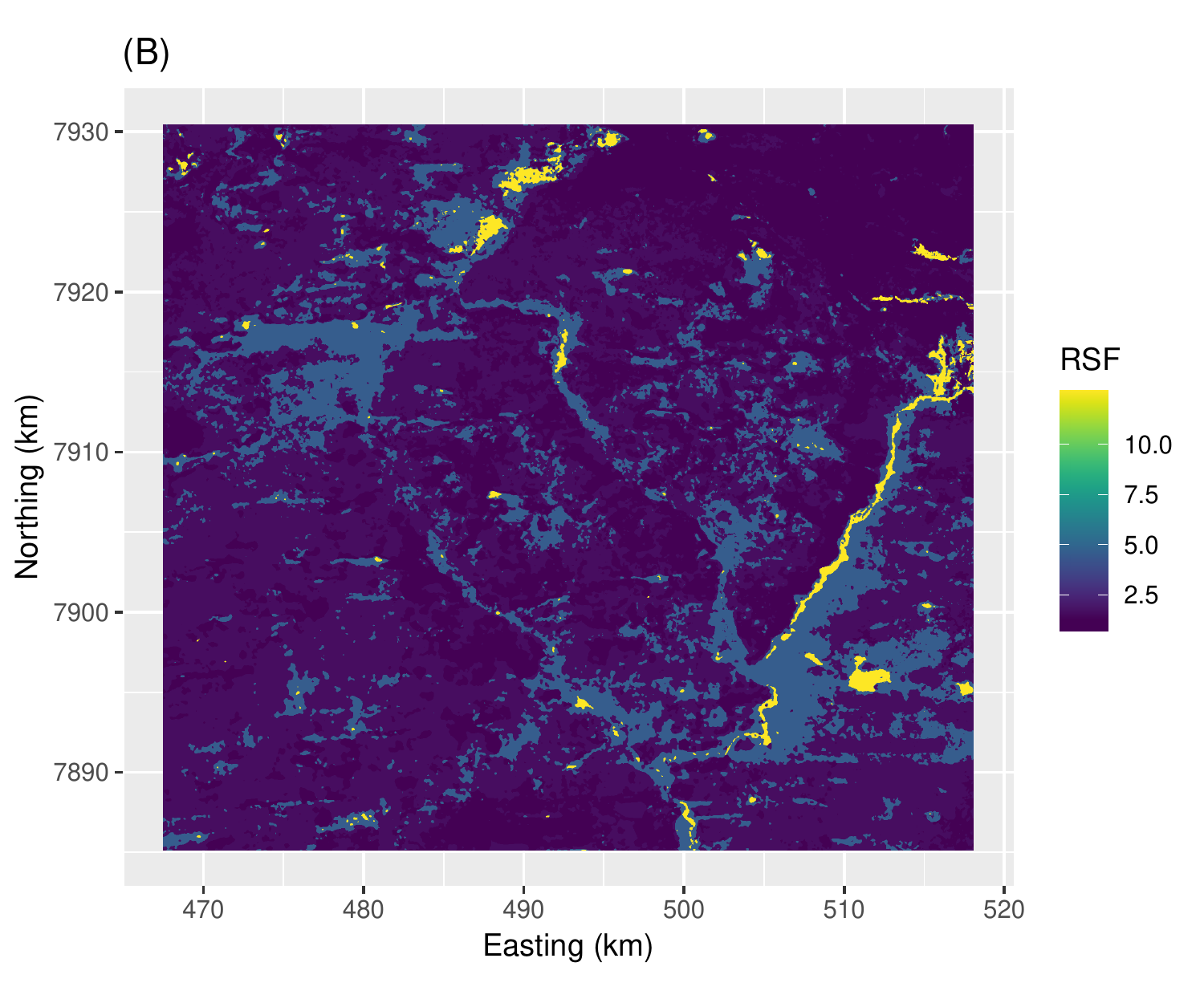}
    \caption{(A) Map of the habitats, with the zebra track overlaid (black line). (B) Estimated RSF in the zebra case study, from the local Gibbs model with normal transition kernel. Colour figures can be viewed in the online version of the paper.}
    \label{fig:zeb1}
\end{figure}

The estimates of the habitat selection parameters and the Hessian-based 95\% confidence intervals are given in Table \ref{tab:zeb1} (under ``Model 1''), and a map of the fitted RSF is shown in Figure \ref{fig:zeb1}(B). The estimated habitat selection parameters indicate that this zebra selects open habitats more strongly than wooded areas, which is consistent with the natural history of the species. Zebras prefer more open areas that provide more forage and greater visibility. This result is also consistent with an analysis based on a standard RSF, conducted by \cite{courbin2016reactive} on many individuals in the same area, albeit with a different vegetation map.

\begin{table}[hbtp]
	\centering
	\begin{tabular}{lccc}
    \toprule
    & Parameter & Model 1 & Model 2 \\
    \midrule
	Grassland & $\hat{\beta}_\text{G}$ & 2.74 (2.54,2.94) & 2.29 (2.03,2.55) \\
    Bushed grassland & $\hat{\beta}_\text{BG}$ & 1.42 (1.25,1.60) & 1.32 (1.08,1.55) \\
    Bushland & $\hat{\beta}_\text{B}$ & 0.01 $(-0.18,0.19)$ & 0.24 (0.01,0.47)\\
    Woodland (reference) & $\hat{\beta}_\text{W}$ & 0 & 0 \\
    \bottomrule
	\end{tabular}
    \caption{Estimates and Hessian-based 95\% confidence intervals of the habitat selection parameters, in the zebra case study, under the local Gibbs model with normal transition density (Model 1), and the local Gibbs model with gamma-distributed availability radius (Model 2). The woodland habitat is the reference category, and the corresponding coefficient is fixed to zero and not estimated.}
    \label{tab:zeb1}
  \end{table}
  
The standard deviation of the transition density was estimated to $\hat\sigma = 0.20$. Under this model, in the absence of covariate effects, the step lengths of the animal follow a Rayleigh distribution with scale parameter $\lambda = \sqrt{2}\sigma$. The estimate of the scale parameter is $\hat\lambda = 0.28$, and the mean of the (resource-indendepent) step length distribution can be derived as $\sqrt{\pi/2} \hat\lambda = 0.35$km. To assess this movement model, we simulated $10^4$ locations from the fitted model, on the same habitat map as the observations. We compared the distribution of step lengths observed in the zebra data set to the distribution of simulated step lengths (Figure \ref{fig:steps}). There is a clear discrepancy between the two distributions: the model fails to capture very short and very long step lengths, and overestimates the density of intermediate step lengths. The empirical distribution of step lengths has a mode at zero, and a long tail, which cannot be appropriately modelled by this formulation. We then considered the random availability radius model for more flexibility.

\subsection{Random availability radius model}
\label{sec:appli2}
We fitted the local Gibbs model with random availability radius, described in Section \ref{sec:model3}, to the same track. We modelled the availability radius with a gamma distribution, and estimated its shape and rate parameters. We used Monte Carlo samples of size $n_r = n_c = n_z = 30$, following the simulation study of Section \ref{sec:sim3}. As in Section \ref{sec:appli1}, we ran the numerical optimisation 50 times with random initial parameter values, and kept the model with the largest likelihood, to avoid numerical convergence issues. Each model fit took about 1.5 hour on a 2GHz i5 CPU. We evaluated the Hessian matrix of the log-likelihood at the maximum likelihood estimates, with $n_r = n_c = n_z = 50$, and derived standard errors for the parameters.

The estimates of the habitat selection parameters, and the 95\% confidence intervals, are given in Table \ref{tab:zeb1} (under ``Model 2''). The parameter values are quite similar to those obtained with the normal kernel model, and the results confirm that the selection is stronger for open habitats (i.e.\ grassland and bushed grassland). The estimated shape of the gamma distribution of the availability radius was $\hat\alpha = 0.77$, and the rate was $\hat\rho = 3.56$. The estimated gamma distribution of the availability radius therefore had mean $\hat{E}(r_t) = \hat\alpha/\hat\rho = 0.22$km, and 95th percentile $\hat{P}_{0.95}=0.71$km.

\begin{figure}[hbtp]
	\centering
    \includegraphics[width=0.7\textwidth]{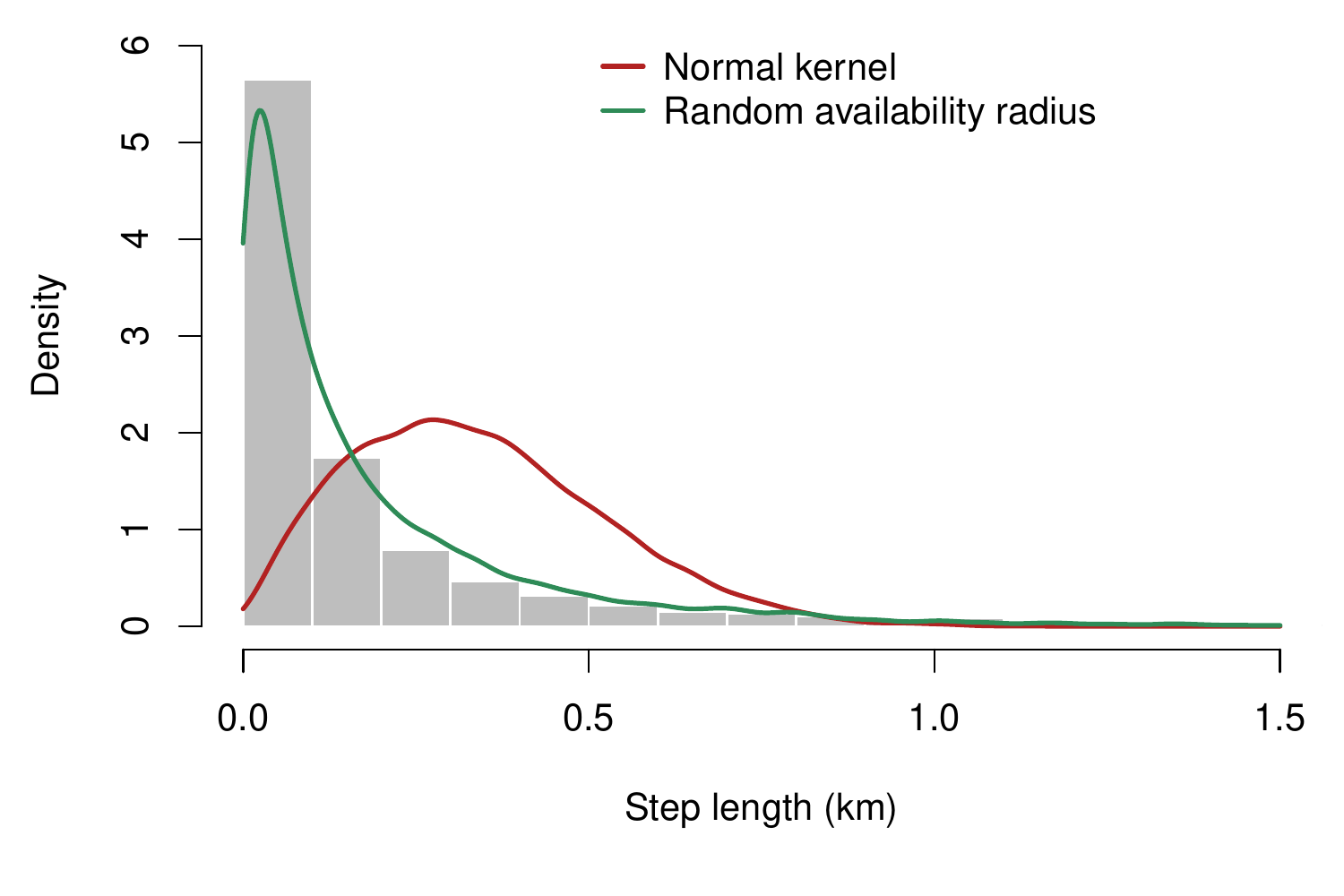}
    \caption{Histogram of the observed step lengths in the zebra data set. The lines show the densities of simulated step lengths, obtained from two fitted models: the local Gibbs model with normal transition density, and the local Gibbs model with gamma-distributed availability radius. We truncated the $x$-axis to $[0,1.5]$ for better visualisation, but the maximum observed step length is around 3km. Colour figures can be viewed in the online version of the paper.}
    \label{fig:steps}
\end{figure}

To assess the random availability radius model, we cannot directly measure the goodness-of-fit for the estimated distribution of the availability radius, because its true value is never observed. Instead, like we did for the normal kernel model in Section \ref{sec:appli1}, we simulated a track of length $10^4$ from the fitted model, on the same habitat map. We compared the distributions of observed and simulated step lengths (Figure \ref{fig:steps}). The distribution of the simulated steps resembles that of the observed steps much more closely than with the normal kernel model. This indicates that the model was able to capture the speed of the zebra's movement. This is remarkable, as the step lengths or the speeds are never directly modelled: instead, we estimated the distribution of the unobserved radius of the relocation region.

There is a trade-off between realism of the movement model and computational speed: the random availability radius model was 15-20 times slower than the normal kernel model in this analysis, due to the additional nested integral in its likelihood (Equation \ref{eqn:RARlike}). Here, the habitat selection estimates were very similar using both models. This suggests that the simpler one (normal kernel model) is sufficient to capture the RSF, even if the movement component is not flexible enough to capture the zebra's step lengths. However, we could not have known this before fitting the random availability radius model and, generally, model checking methods should be used to verify that features of the movement are appropriately captured by the model.

\section{Discussion}
We showed how a new class of step selection models, based on the simulation rules of MCMC algorithms, can be used to estimate an animal's habitat selection and movement characteristics. We provided a natural framework to model animal movement and space use, in which short-term step selection gives rise to the long-term utilisation distribution. This approach connects standard RSF and SSF models, as the equilibrium distribution of the movement model is guaranteed to be proportional to the underlying resource selection function. We described maximum likelihood estimation for the local Gibbs sampler, a flexible family of MCMC algorithms which can be used to model animal movement. Parameters of movement and habitat selection can be estimated jointly.

Although the simulations and the case study were based on categorical habitat data, the method would also be applicable to continuous environmental covariates, such as elevation or distance to water. One could for example model territoriality or attraction towards a point in space, with the inclusion of the distance to that point as a covariate.

In the case study of Section \ref{sec:appli}, we fitted two different local Gibbs models to a zebra movement track, and compared them using simulations. We found that simulated step lengths from the local Gibbs model with random availability radius matched the observed step lengths more closely than those simulated from the local Gibbs model with normal transition density. However, the computational effort was greatly increased in the former, because of the need to integrate over the random availability radius. It may be possible to find a local Gibbs formulation that combines the computational speed of the normal kernel model and the flexibility of the random availability radius model. For example, we could define the transition density ($\phi$ in Section \ref{sec:model2}) as the combination of a uniform distribution of angles and a given distribution for the distance to the origin. The uniform angles ensure that the transition density is symmetric, and the shape of the distance distribution determines the resource-independent movement model. It may be possible to achieve a distribution of step lengths with a mode close to zero, as in the zebra data set, with an exponential or Weibull distribution of distances.

An important feature of the local Gibbs model is that the size of the region of availability does not need to be defined a priori. In habitat selection analyses based on use-availability designs, the choice of the spatial extent of the availability region is challenging, and can lead to biased selection estimates \citep{beyer2010interpretation, northrup2013practical}. Instead, we estimate it from the observed tracking data, with a movement model based on a symmetric transition density. The scale of availability is for example measured by the variance of the normal kernel model (Section \ref{sec:model3}), and by the radius in the availability radius model (Section \ref{sec:model4}). One limitation of this method is that the animal's movement and perception are modelled jointly. The transition density of the algorithm describes both the size of the region that the animal can perceive and the distance that it is susceptible to cover over one time interval. This is a strong assumption, that is made in virtually all step selection models \citep{fortin2005wolves, forester2009accounting}, in which habitat selection is considered to take place at the scale of the movement kernel. To the best of our knowledge, only \cite{avgar2015space} have built a step selection model which estimates the movement process and the perception separately. Additional work is required to allow this flexibility within the framework presented in this paper.

\subsection*{Acknowledgements}
TM was supported by the Centre for Advanced Biological Modelling at the University of Sheffield, funded by the Leverhulme Trust, award number DS-2014-081. SCJ was supported by the grant ANR-16-CE02-0001-01 of the French \emph{Agence Nationale de la Recherche}.

\bibliographystyle{apalike}
\bibliography{refs.bib}

\renewcommand{\thefigure}{A\arabic{figure}}
\setcounter{figure}{0}
\newpage
\section*{Appendix 1: Implementation of the approximate local Gibbs likelihood}
\subsection*{Normal kernel model}
The approximate likelihood of a step from $\bm{x}_t = (x_t,y_t)$ to $\bm{x}_{t+1} = (x_{t+1},y_{t+1})$, under the normal kernel model, can be calculated as
\begin{equation*}
	\hat{p}(\bm{x}_{t+1} \vert \bm{x}_t, \bm{\beta}, \sigma) = 
    	\pi(\bm{x}_{t+1} \vert \bm\beta) \dfrac{n_z}{n_c}
    	\sum_{i=1}^{n_c} \dfrac{\varphi(\bm{x}_{t+1} \vert \bm\mu_i, \sigma^2 \bm{I}_2)}
        {\sum_{j=1}^{n_z} \pi(\bm{z}_{ij} \vert \bm\beta)},
\end{equation*}
where 
\begin{itemize}
	\item $\forall i \in \{ 1, \dots, n_c \}, \bm\mu_i \sim N(\bm{x}_t, \sigma^2 \bm{I}_2)$,
    \item $\forall j \in \{ 1, \dots, n_z \}, \bm{z}_{ij} \sim N(\bm\mu_i, \sigma^2 \bm{I}_2)$.
\end{itemize}

Standard random generation functions, such as \texttt{rnorm} in R, can be used to obtain the Monte Carlo samples. Alternatively, uniform samples can be generated (e.g.\ to make use of Latin hypercube sampling), and transformed through the quantile function of the normal distribution.

\subsection*{Random availability radius model}
The approximate likelihood of a step from $\bm{x}_t = (x_t,y_t)$ to $\bm{x}_{t+1} = (x_{t+1},y_{t+1})$, under the random availability radius model, can be calculated as
\begin{equation*}
	\hat{p}(\bm{x}_{t+1} \vert \bm{x}_t, \bm{\beta}, \bm\omega) = 
    \dfrac{\pi(\bm{x}_{t+1} \vert \bm\beta)}{\pi^2} \dfrac{n_z}{n_r n_c}
    \sum_{i=1}^{n_r} \dfrac{A(\mathcal{D}_{r_i}(\bm{x}_t) \cap \mathcal{D}_{r_i}(\bm{x}_{t+1}))}{r_i^4}  
    \sum_{j=1}^{n_c} \dfrac{1}{\sum_{k=1}^{n_z} \pi(\bm{z}_{ijk} \vert \bm\beta)},
\end{equation*}
where $A(\cdot)$ denotes the area, and
\begin{itemize}
	\item $\forall i \in \{ 1, \dots, n_r \}, r_i \sim p(r \vert \bm\omega)$,
    \item $\forall j \in \{ 1, \dots, n_c \}, \bm\mu_{ij} \sim U(\mathcal{D}_{r_i}(\bm{x}_t) \cap \mathcal{D}_{r_i}(\bm{x}_{t+1}))$,
    \item $\forall k \in \{ 1, \dots, n_z \}, \bm{z}_{ijk} \sim U(\mathcal{D}_{r_i}(\bm\mu_{ij}))$.
\end{itemize}

The implementation thus relies on the generation of three sets of Monte Carlo samples, for each observed step: the availability radii $\{r_i\}$, the intermediate centres $\{\bm\mu_{ij}\}$, and the end points $\{\bm{z}_{ijk}\}$.

\paragraph{The availability radii} The $r_i$ are drawn from the probability density $p(r \vert \bm{\omega})$. Note that sampled radii that are smaller than $d_t/2$ do not contribute to the likelihood, where $d_t$ is the distance between $\bm{x}_t$ and $\bm{x}_{t+1}$. Indeed, $A(\mathcal{D}_{r_i}(\bm{x}_t) \cap \mathcal{D}_{r_i}(\bm{x}_{t+1}))=0$ when $r_i \leq d_t/2$. It is then more economical to sample the $r_i$ from $p(r \vert \bm{\omega})$ truncated to $[d_t/2,\infty)$, and to multiply the likelihood by $1-F_r(d_t/2 \vert \bm{\omega})$, where $F_r$ is the cumulative distribution function of the radius.

One way to sample from the truncated distribution is to generate a uniform sample $u_i \sim U(0,1)$, and transform it with the quantile function $\tilde{F}_r$ of the truncated distribution:
\begin{equation*}
	r_i = \tilde{F}_r^{-1}(u_i) = F_r^{-1} [ F_r(d_t/2) + u_i (1 - F_r(d_t/2)) ].
\end{equation*}
See \cite{nadarajah2006r} for details on the implementation of truncated distributions.

\begin{figure}[htbp]
	\centering
	\includegraphics[width=0.5\textwidth]{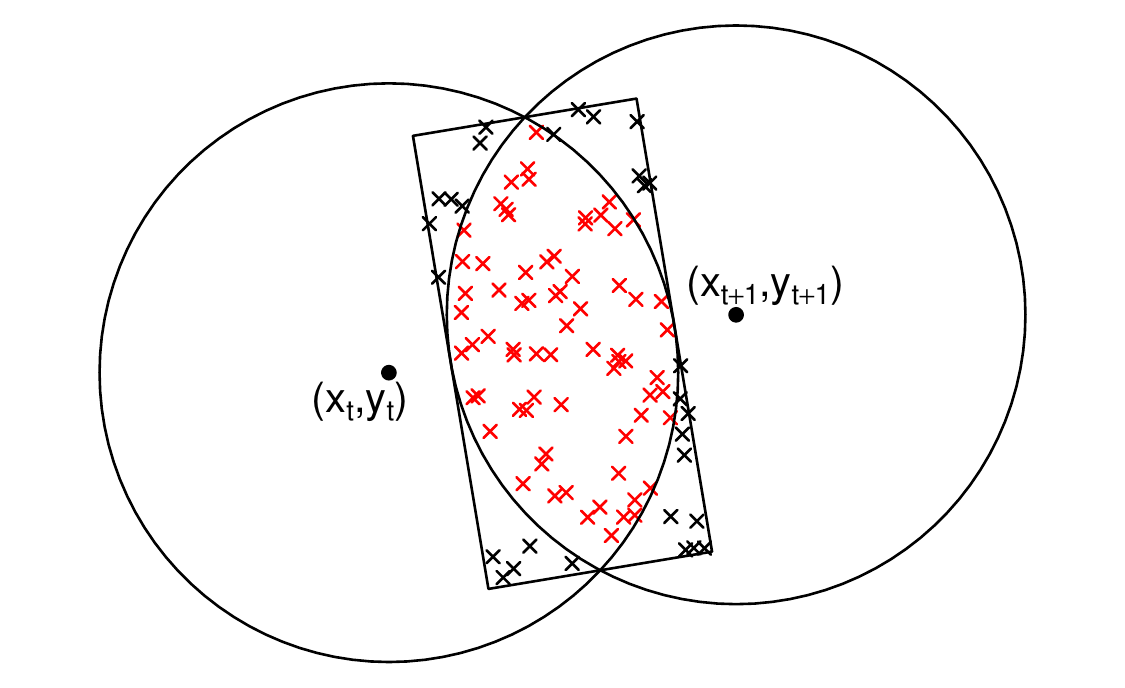}
    \caption{Simulation of points in the intersection of two discs. The crosses are sampled uniformly from the rectangle, and only the red crosses are accepted.}
    \label{fig:discs}
\end{figure}

\paragraph{The intermediate centres} For each $r_i$, the intermediate points $\bm\mu_{ij}$ are sampled uniformly from the intersection of the discs $\mathcal{D}_{r_i}(\bm{x}_t)$ and $\mathcal{D}_{r_i}(\bm{x}_{t+1})$. In practice, one way to do this is to generate points uniformly from the smallest rectangle which includes the intersection of the discs, illustrated in Figure \ref{fig:discs}, and to reject those which do not lie within a distance $r_i$ of both $\bm{x}_t$ and $\bm{x}_{t+1}$.

A point can be sampled uniformly from the intersection of the discs as follows. Sample a point $(u_x,u_y)$ uniformly from the square of side 1 and center (0,0):
\begin{equation*}
	\begin{cases}
		u_x \sim U(-0.5,0.5) \\
        u_y \sim U(-0.5,0.5)
	\end{cases}
\end{equation*}

Then, scale both dimensions onto a rectangle of correct width and height:
\begin{equation*}
	\begin{cases}
		u'_x = (2r_i-d_t) u_x \\
        u'_y = (2 \sqrt{r_i^2 - d_t^2/4}) u_y
	\end{cases}
\end{equation*}

Move to polar coordinates, to rotate the rectangle:
\begin{equation*}
	\begin{cases}
		l = \sqrt{(u'_x)^2 + (u'_y)^2} \\
        \theta = \text{atan2} (u'_y, u'_x) 
	\end{cases}
\end{equation*}

Rotate the rectangle, to align it with the step from $\bm{x}_t$ to $\bm{x}_{t+1}$:
\begin{equation*}
	\theta' = \theta + \text{atan2}(y_{t+1}-y_t, x_{t+1}-x_t)
\end{equation*}

Finally, convert the sample back to cartesian coordinates, and translate the centre of the rectangle to the mid-point between $\bm{x}_t$ and $\bm{x}_{t+1}$.
\begin{equation*}
	\begin{cases}
		u''_x = l \cos(\theta') + (x_t + x_{t+1})/2 \\
        u''_y = l \sin(\theta') + (y_t + y_{t+1})/2
	\end{cases}
\end{equation*}

The point $(u''_x,u''_y)$ is sampled uniformly from the rectangle shown in Figure \ref{fig:discs}. Accept it (i.e.\ $\bm\mu_{ij} = (u''_x,u''_y)$) if it is in $\mathcal{D}_{r_i}(\bm{x}_t) \cap \mathcal{D}_{r_i}(\bm{x}_{t+1})$, else reject it. Repeat until $n_c$ points have been accepted.

\paragraph{The end points} For each radius $r_i$ and centre $\bm\mu_{ij}$, the locations $\bm{z}_{ijk}$ are sampled uniformly from the disc $\mathcal{D}_{r_i}(\bm\mu_{ij})$.

For each $k = 1,\dots,n_z$, take
\begin{equation*}
	\begin{cases}
		l_k \sim U(0,r_i^2) \\
        \theta_k \sim U(-\pi,\pi)
	\end{cases}
\end{equation*}

Then,
\begin{equation*}
	\bm{z}_{ijk} = \bm\mu_{ij} + \sqrt{l_k}
    \begin{pmatrix}
    	\cos \theta_k \\
        \sin \theta_k
    \end{pmatrix}
\end{equation*}
is uniformly sampled from $\mathcal{D}_{r_i}(\bm\mu_{ij})$.

\end{document}